\documentclass[preprint]{aastex}

\newcommand{\stateone}{B component}
\newcommand{\statetwo}{D component}

\slugcomment{Accepted for publication in the Astrophysical Journal}

\shorttitle{Structural Variation of Molecular Gas}
\shortauthors{Sawada et al.}

\begin{document}


\title{Structural Variation of Molecular Gas in the Sagittarius Arm
  and Inter-Arm Regions}

\author{Tsuyoshi Sawada\altaffilmark{1,2},
  Tetsuo Hasegawa\altaffilmark{1},
  and Masahiro Sugimoto\altaffilmark{1}}
\affil{Joint ALMA Office, Alonso de C\'{o}rdova 3107, Vitacura,
  Santiago 763-0355, Chile}
\email{sawada.tsuyoshi@nao.ac.jp}

\author{Jin Koda}
\affil{Department of Physics and Astronomy,
  Stony Brook University, Stony Brook, NY 11794-3800}

\and

\author{Toshihiro Handa\altaffilmark{3}}
\affil{Department of Physics, Faculty of Science, Kagoshima University,
  1-21-35 Korimoto, Kagoshima, Kagoshima 890-0065, Japan}

\altaffiltext{1}{NAOJ Chile Observatory,
  Joaqu\'{\i}n Montero 3000 Oficina 702, Vitacura,
  Santiago 763-0409, Chile}
\altaffiltext{2}{Nobeyama Radio Observatory, 462-2 Nobeyama, Minamimaki,
  Minamisaku, Nagano 384-1305, Japan}
\altaffiltext{3}{Institute of Astronomy, The University of Tokyo,
  2-21-1 Osawa, Mitaka, Tokyo 181-0015, Japan}


\begin{abstract}
We have carried out survey observations toward the Galactic plane
at $l\approx 38\degr$ in the $^{12}{\rm CO}$ and $^{13}{\rm CO}$
$J=1\mbox{--}0$ lines using the Nobeyama Radio Observatory 45-m
telescope.
A wide area ($0\fdg 8\times 0\fdg 8$) was mapped with
high spatial resolution ($17\arcsec$).
The line of sight samples the gas in both the Sagittarius arm and
the inter-arm regions.
The present observations reveal
how the structure and physical conditions vary across a spiral arm.
We classify the molecular gas in the line of sight
into two distinct components based on its appearance:
the bright and compact {\it \stateone}
and the fainter and diffuse (i.e., more extended) {\it \statetwo}.
The {\stateone} is predominantly seen at the spiral arm velocities,
while the {\statetwo} dominates at the inter-arm velocities
and is also found at the spiral arm velocities.
We introduce the {\it brightness distribution function} and
the {\it brightness distribution index} (BDI, which indicates
the dominance of the {\stateone})
in order to quantify the map's appearance.
The radial velocities of BDI peaks coincide with those of high
$^{12}{\rm CO}\;J=3\mbox{--}2/^{12}{\rm CO}\;J=1\mbox{--}0$
intensity ratio (i.e., warm gas) and \ion{H}{2} regions,
and tend to be offset from the line brightness peaks
at lower velocities (i.e., presumably downstream side of the arm).
Our observations reveal that the gas structure at small scales
changes across a spiral arm:
bright and spatially confined structures develop in a spiral arm,
leading to star formation at downstream side,
while extended emission dominates in the inter-arm region.
\end{abstract}




\section{Introduction}\label{sec:introduction}

The interstellar medium (ISM) plays an important role in galaxies:
stars, the principal constituent of galaxies,
are born from and return matter to the ISM.
Since most neutral ISM is molecular in the inner part of galaxies
\citep[e.g.,][]{dame1993,sofue1995,honma1995,koda2009}
and stars form in dense molecular clouds,
the studies of the distribution, spatial structures, and physical
conditions of molecular gas are essential to understand the star-forming
activities in galaxies.
The Milky Way Galaxy is the unique target for resolving
sub-pc structures of molecular gas with radio telescopes
due to its proximity.

Some extensive mapping surveys of the Galactic disk in CO emission lines
have been made since \citet{scoville1975} and \citet{gordon1976}.
Two historic $^{12}{\rm CO}\;J=1\mbox{--}0$ surveys are the Columbia-CfA
survey \citep[][and references therein]{dame1987,dame2001} and
the Massachusetts-Stony Brook Galactic Plane CO Survey
\citep{sanders1986,clemens1986}.
These surveys made use of the Columbia-CfA 1.2-m and the Five College
Radio Astronomy Observatory (FCRAO) 14-m telescopes, providing an
angular resolutions of $8\arcmin$ and $45\arcsec$,
respectively\footnote{The Massachusetts-Stony Brook survey data were
under\-sampled at $3\arcmin$--$6\arcmin$, while the Columbia-CfA data
were mostly full-beam ($7\farcm 5$) sampled.}.
The first Galactic quadrant (and portions of adjacent quadrants)
was mapped in the optically thinner isotopologue
$^{13}{\rm CO}\;J=1\mbox{--}0$ line
by the Bell Laboratories survey \citep{lee2001} with their 7-m telescope
($103\arcsec$ resolution) and 
the Boston University-FCRAO Galactic Ring Survey \citep{jackson2006}
with the 14-m telescope.
Surveys in higher-$J$ transitions \citep[e.g.,
$^{12}{\rm CO}\;J=2\mbox{--}1$ by][]{sakamoto1995,sakamoto1997}
were also made to diagnose the physical conditions of the gas.

These surveys have revealed the basic properties of molecular gas, such
as its large-scale distribution in the Galaxy
\citep{dame1986,clemens1988}, and the presence of discrete molecular
entities, i.e., giant molecular clouds \citep{solomon1979}.
There were many studies focused on identifying discrete molecular clouds
and determining their physical/statistical properties
\citep[e.g., mass spectrum and the size-line\-width
relation:][]{sanders1985,solomon1987}
and of sub\-structures within them \citep[e.g.,][]{simon2001}.
\citet{solomon1985,sanders1985,scoville1987} studied the distribution
of molecular clouds in the first Galactic quadrant.
Although the catalogued clouds were distributed rather uniformly
in the longitude-velocity ($l$-$v$) plane,
they found that a subset of the clouds, ``warm'' or ``hot'' clouds
with high brightness temperatures, followed clear $l$-$v$ patterns.
The patterns agreed with those traced by \ion{H}{2} regions and
were considered as spiral arms.
This might reflect the variation of the characteristics of the clouds
affected by the galactic structures.

Previous surveys employed sparse spatial sampling and/or low resolution
(on the order of arc\-minutes).
Recently developed instruments have enabled us
to perform Nyquist-sampled, high-resolution (tens of arc\-seconds)
survey observations over wide fields \citep[e.g.,][]{jackson2006}.
With the high quality data, we would like to re-evaluate
the picture of molecular content in the Galaxy.
In particular, is there a way to study the spatial structures
and the properties of the gas without decomposing it
into clouds or their sub\-structures, since the decomposition
inevitably omits a part of the emission which may be important.

In this paper we present the results from our observations
in the $^{12}{\rm CO}$ and $^{13}{\rm CO}$ $J=1\mbox{--}0$ lines
with the Nobeyama Radio Observatory (NRO) 45-m telescope,
proposing an alternative, complementary method to study
the spatial structure of the gas at a sub-pc resolution.
The method is based on the histogram of the brightness temperature
of the line emission, or the {\it brightness distribution function}.
We show how we selected the field, at $l\approx 38\degr$,
in Section \ref{sec:field}.
The observations and the data reduction process are described in
Section \ref{sec:observations}.
In Section \ref{sec:results} the velocity channel maps are presented.
Then we characterize the observed
line brightness, and discuss how the characteristics of the gas
change across the spiral arms.


\section{Field Selection}\label{sec:field}

\begin{figure}
\epsscale{0.8}
\plotone{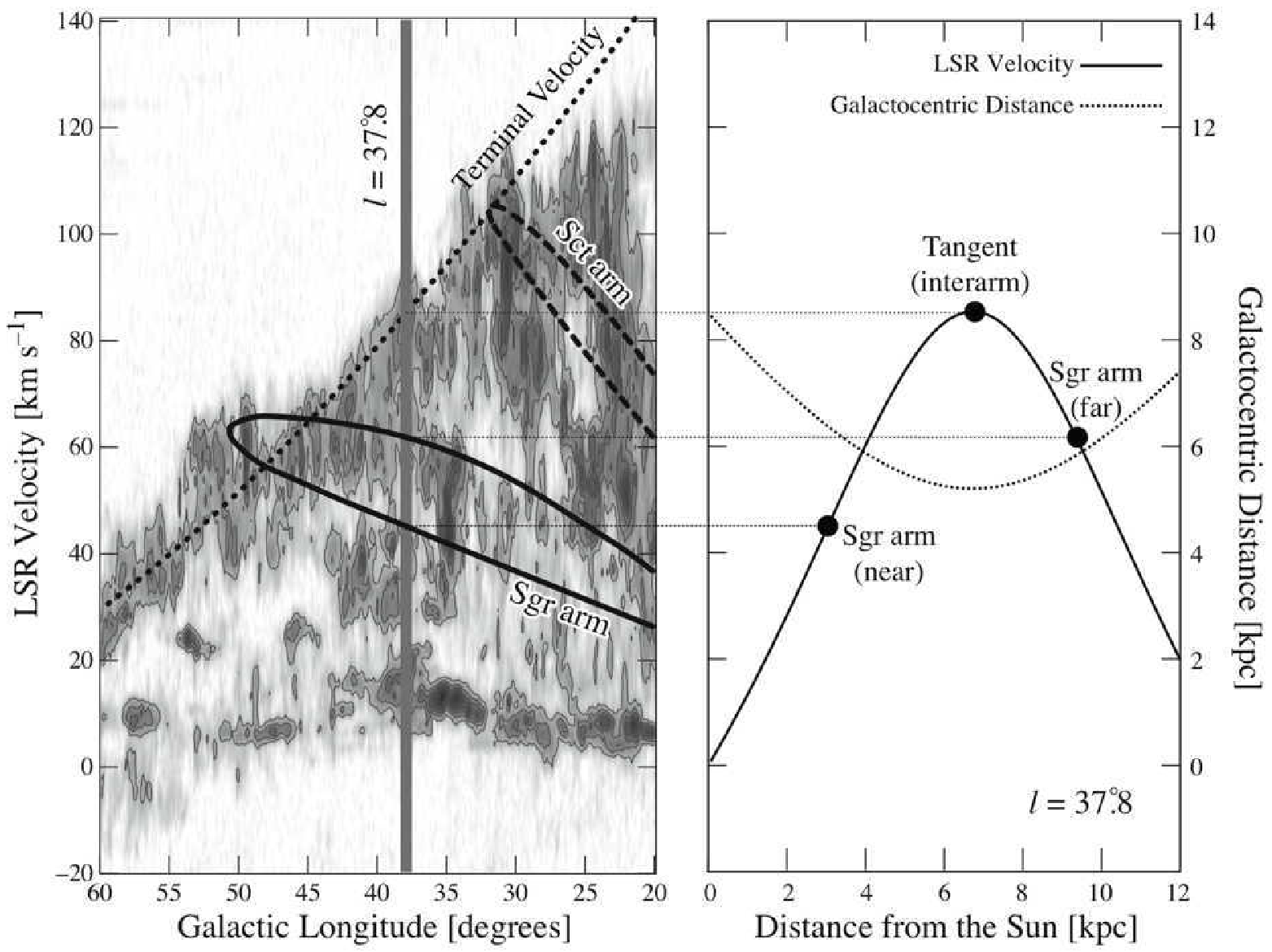}
\caption{({\it a\/})
  Loci of the Sgr arm ({\it solid line\/}) and
  the Sct arm ({\it dashed line\/}) by \citet{sanders1985}
  overlaid on the $^{12}{\rm CO}\; J=1\mbox{--}0$ longitude-velocity
  diagram \citep{dame2001}.
  The tangent velocity for
  the $220\;{\rm km\; s^{-1}}$ flat rotation is also shown
  ({\it dotted line\/}).
  The observed line of sight is shaded.
  ({\it b\/})
  LSR velocity ({\it solid line\/}; left ticks)
  and Galacto\-centric distance ({\it dashed line\/}; right ticks)
  as functions of distance from the Sun toward $l = 37\fdg 8$.
  \label{fig:lvpunch}}
\end{figure}

Fig.\ \ref{fig:lvpunch}a shows the longitude-velocity diagram of the
$^{12}{\rm CO}\; J=1\mbox{--}0$ line \citep{dame2001}.
The loci of the Sagittarius and Scutum arms \citep{sanders1985} and the
tangent velocity\footnote{Throughout the paper, we use the distance from
the Galactic center to the Sun of 8.5 kpc and assume the
220-${\rm km\;s^{-1}}$ flat rotation of the Milky Way Galaxy.}
are overlaid.

We have chosen a field centered at $l=37\fdg 8$ for the following
reason.
This longitude is located between
the Sct tangent ($l \simeq 30\degr$) and
the Sgr tangent ($l \simeq 50\degr$), and therefore
it intersects the Sgr arm twice.
According to the model of \citet{sanders1985}, the radial velocities of
the intersections are $v_{\rm LSR} \simeq 45$ and
$60\;{\rm km\; s^{-1}}$.
With the flat, circular rotation of the Galaxy, these velocities
correspond to the near- and far-side intersections, respectively.
This is consistent with the distances to molecular clouds determined by
\citet{solomon1987}.
Around this longitude
the clouds at $\simeq 40\;{\rm km\;s^{-1}}$ are mostly on the near side,
while those at $\simeq 60\;{\rm km\;s^{-1}}$ are on the far side.
It should be noted that inter-arm emission at the opposite sides
potentially contaminates that from the two arm intersections.
On the other hand, the tangent velocity
($v_{\rm LSR}\simeq 85\;{\rm km\; s^{-1}}$) component samples the
inter-arm region (the tangent point)
between the Sct and Sgr arms, without the distance ambiguity.
The near- and far-side Sgr arm and the tangent component are all
sufficiently bright in the CO line (Fig.\ \ref{fig:lvpunch}a) and are
separated from each other by $\simeq 20\;{\rm km\; s^{-1}}$,
which allows us to compare the properties of arm and inter-arm emission
within a single field of view.

The radial velocity and the
Galacto\-centric distance as a function of distance from the observer
are shown in Fig.\ \ref{fig:lvpunch}b.
The distances to the near- and far-sides of the Sgr arm, and the
inter-arm region (tangent velocity) are $\simeq 3$, 9, and
$6.7\;{\rm kpc}$, respectively.
The ratio between the far and near kinematic distances for the 45-
and 60-${\rm km\; s^{-1}}$ components are 3.4 and 2.3, respectively:
i.e., a misidentification of the distance causes an over- or
under-estimation of the linear scale of structures of the gas
by a factor of a few (2.3--3.4).
The main outcome of this paper is not affected by this uncertainty,
as discussed in Section \ref{subsec:bdf}.
Possible deviation from flat, circular rotation
due to streaming motion and/or random motion
causes errors in the estimated kinematic distances.
A velocity shift of $\pm 10\;{\rm km\; s^{-1}}$ results in a distance
error of $\pm 0.6$ kpc (20\%), $\pm 0.7$ kpc (7\%), and $\pm 1.6$ kpc
(24\%) at the near- and far-sides of the Sgr arm, and the tangent
component, respectively.
The Galacto\-centric distance of these components is typically 6 kpc.

Although we assumed the structures of the Galaxy as described above,
there is uncertainty of, e.g., the distribution and location of the
arms.
Possible deviation from the assumed geometry and its influence on
the discussion are described in Appendix \ref{app:sgrarm}.


\section{Observations and Data Reduction}\label{sec:observations}


\subsection{CO $J=1\mbox{--}0$ Observations}

Observations of the $^{12}{\rm CO}\; J=1\mbox{--}0$ (rest frequency of
115.271 GHz) and $^{13}{\rm CO}\; J=1\mbox{--}0$ (110.201 GHz)
transitions were carried out in 2002--2003 (Period 1) and 2005--2006
(Period 2) using the NRO 45-m telescope.
The observing parameters are summarized in Table \ref{tab:obs}.
The half-power beam width (HPBW) of the telescope at 115 GHz was
$15\arcsec$.
At the distances of the near- and far-side Sgr arm and the tangent
point, this corresponds to 0.22, 0.65, and 0.49 pc, respectively.
The main-beam efficiency ($\eta_{\rm MB}$) was
$\simeq 0.4$ (see Table \ref{tab:obs}).
The forward spillover and scattering efficiency \citep{kutner1981}
at 115 GHz was measured using the moon in November 2003,
$\eta_{\rm moon} = 0.69\pm 0.03$.

We used the 25-BEam Array Receiver System \citep[BEARS:][]{sunada2000}.
The front\-end consists of a $5\times 5$ focal-plane array of
double-sideband (DSB),
superconductor-insulator-superconductor (SIS) mixer receivers
with a beam separation of $41\farcs 1$ \citep{yamaguchi2000}.
We adopted the chopper-wheel method,
switching between a room-temperature load and the sky,
for primary intensity calibrations.
This corrects for atmospheric attenuation and antenna losses,
and converts the intensity scale
to the antenna temperature in DSB [$T_{\rm A}^*({\rm DSB})$].
The back\-end was a set of 1024-channel digital auto\-correlation
spectrometers \citep{sorai2000}.
It was used in the wide-band and high-resolution modes
(band\-width of 512 and 32 MHz, respectively: Table \ref{tab:obs}).

In Period 1, we mapped the $^{13}{\rm CO}$ line in a
$0\fdg 3 \times 0\fdg 5$ region
($37\fdg 43 \lesssim l \lesssim 37\fdg 80$,
$-0\fdg 52 \lesssim b \lesssim +0\fdg 02$)
using the position-switch (PSW) observing method.
The spectrometer was mainly used in the high-resolution mode.
The grid spacing was chosen to be $13\farcs 7$,
one third of the beam separation of BEARS.
An emission-free reference ({\sc off}) position was taken at
$(l,b) \approx (37\fdg 5,-1\fdg 5)$ for three
20-s on-source integrations.
The total number of observed points was about 13500, and typical
integration time per point was 80 s.
We also mapped the same area at a sparse spatial sampling
using the wide-band mode to determine the spectral baseline range.

In Period 2, we employed the On-The-Fly (OTF) mapping technique
\citep{sawada2008} to map the $^{12}{\rm CO}$ and $^{13}{\rm CO}$ lines
in a $0\fdg 8 \times 0\fdg 8$ region
($37\fdg 35 \le l \le 38\fdg 15, -0\fdg 50 \le b \le +0\fdg 30$).
The size of the map ($0\fdg 8$) corresponds to 40, 130, and 90 pc at the
distances of the near- and far-side Sgr arm and the tangent point,
respectively.
The wide-band mode of the spectrometer was used to observe the
$^{12}{\rm CO}$ line, while the $^{13}{\rm CO}$ data were taken with
both modes.
The sampling interval along the scan rows was set to be $3\arcsec$, and
the separation between the scan rows was $5\arcsec$.
An {\sc off} position was observed before every scan, whose duration was
typically 40 s.
The {\sc off} position $(l,b) \approx (38\fdg 0,-1\fdg 7)$ had been
confirmed to be emission-free in $^{12}{\rm CO}\; J=1\mbox{--}0$:
i.e., $< 0.07\;{\rm K}$ in $T_{\rm A}^*({\rm DSB})$,
which corresponds to $\lesssim 0.15\;{\rm K}$ in single sideband (SSB).
Scans were made in two orthogonal directions, i.e., along $l$ and $b$,
in order to minimize scanning artifact (systematic errors along the
scan direction) in the data reduction process (see Section
\ref{sec:reduction}).

The pointing of the telescope was calibrated by observing SiO
$J=1\mbox{--}0$ maser (42.821 and 43.122 GHz) sources OH39.7+1.5 and
R Aql with another SIS receiver at 40-GHz band (S40) every 1--2 hours.
The pointing accuracy was typically $6\arcsec$ (Period 1) and
$7\arcsec$ (Period 2).

Since BEARS is a DSB receiver system, we need to correct the observed
intensity scale to that of an SSB receiver.
The scaling factors to convert
$T_{\rm A}^*({\rm DSB})$ to $T_{\rm A}^*({\rm SSB})$
were derived by observing a calibrator with S100, the single-beam
SIS receiver equipped with an SSB filter, and with every beam of BEARS.
We used the factors provided by the observatory in Period 1,
and measured them ourselves by observing W51 in Period 2.
Based on multiple measurements,
the reproducibility of the factors in Period 2 was
4\% ($^{12}{\rm CO}$) and 9\% ($^{13}{\rm CO}$).


\subsection{CO $J=1\mbox{--}0$ Data Reduction}\label{sec:reduction}

The $^{\rm 13}{\rm CO}$ PSW data were reduced with the {\it NEWSTAR}
reduction package developed at NRO \citep{ikeda2001}.
During the data reduction, it turned out that there was weak emission
of $T_{\rm A}^* \simeq 0.3\;{\rm K}$ and
$v_{\rm LSR} \simeq 42\;{\rm km\;s^{-1}}$ at the {\sc off} position
for the outer 4 receiver beams.
We successfully corrected this error for 3 out of the 4 beams
by removing Gaussian profiles from the {\sc off} data.
For the remaining 1 beam, we could not remove the influence;
therefore we have excluded the beam from the further reduction process.
Due to an aberration originating in the beam-transfer optics of the
telescope, the beams of BEARS were not exactly aligned on a regular
$41\farcs 1$ grid on the sky.
Therefore, after flagging out bad data, the data were resampled onto a
$6\farcs 85$ grid using a convolution with a Gaussian whose full width
at half maximum (FWHM) was $13\farcs 7$.
As a result the effective spatial resolution became $20\arcsec$.
Linear baselines were fitted and subtracted.
Baseline ranges were taken at $v_{\rm LSR} \simeq 30$ and
$100\;{\rm km\;s^{-1}}$, which had been confirmed to be emission-free
using the wide-band data.
Resultant spectra have typical root-mean-square (RMS) noise of 0.4 K at
a resolution of
$20\arcsec \times 20\arcsec \times 0.10\;{\rm km\;s^{-1}}$.

The reduction of the OTF data was made with the {\it NOSTAR} reduction
package \citep{sawada2008}.
Bad data were flagged, and linear (high-resolution) or parabolic
(wide-band) baselines were fitted and subtracted.
The data were mapped onto a square grid of $6\arcsec$ separation
by spatial convolution with a Gaussian-tapered Jinc function
$J_1(\pi r/a)/(\pi r/a) \cdot \exp[-(r/b)^2]$,
where $J_1$ is the first-order Bessel function,
$a=1.55$, $b=2.52$, and $r$ is the distance from the grid point
to the observed position divided by the grid spacing
\citep{mangum2007}.
As a result the effective spatial resolution was $17\arcsec$.
The scanning artifact was reduced by combining the maps made from the
longitudinal and latitudinal scans using the {\it PLAIT} method
\citep{emerson1988}.
Since the $^{13}{\rm CO}$ high-resolution data do not cover sufficient
emission-free velocity ranges for baseline subtraction,
we used the wide-band data as a
reference to determine the spectral baseline.
That is, the difference between the high-resolution and wide-band
spectra was linearly fitted for each spatial grid, and the regression
was added to the high-resolution data.
The final data\-set has an RMS noise level of 0.18 K ($^{12}{\rm CO}$
wide-band), 0.074 K ($^{13}{\rm CO}$ wide-band), and 0.29 K
($^{13}{\rm CO}$ high-resolution), at a resolution of
$17\arcsec \times 17\arcsec \times 2.6 \;{\rm km\;s^{-1}}$ (wide-band)
and $17\arcsec \times 17\arcsec \times 0.2 \;{\rm km\;s^{-1}}$
(high-resolution).

We checked the relative intensity calibration between Periods 1 and 2.
The Period 1 map was resampled onto the same grid as the Period 2 map,
and the correlation of the intensities was examined in the region
in which the maps overlap.
The results are consistent and we found the relation
$T_{\rm A}^*(\mbox{Period 1}) =
(1.048\pm 0.001)\cdot T_{\rm A}^*(\mbox{Period 2})$
for the pixels which were brighter than 1 K in both maps.
The difference of $\approx 5\%$ is within the uncertainties of
$\eta_{\rm MB}$ and the scaling factors converting
$T_{\rm A}^*({\rm DSB})$ into $T_{\rm A}^*({\rm SSB})$.
Thus we combined the maps from two periods.

Hereafter, the line intensities are shown in the main-beam temperature
[$T_{\rm MB} \equiv T_{\rm A}^*({\rm SSB})/\eta_{\rm MB}$] scale
(for the $^{13}{\rm CO}$ data, we adopted the value in Period 2, i.e.,
$\eta_{\rm MB}=0.45$).
The $T_{\rm MB}$ is appropriate for the brightness temperature of
compact structures, whereas the brightness of spatially extended
emission may be overestimated (Appendix \ref{app:efficiency}).


\subsection{CO $J=3\mbox{--}2$ Data}

\begin{figure}
\epsscale{0.8}
\plotone{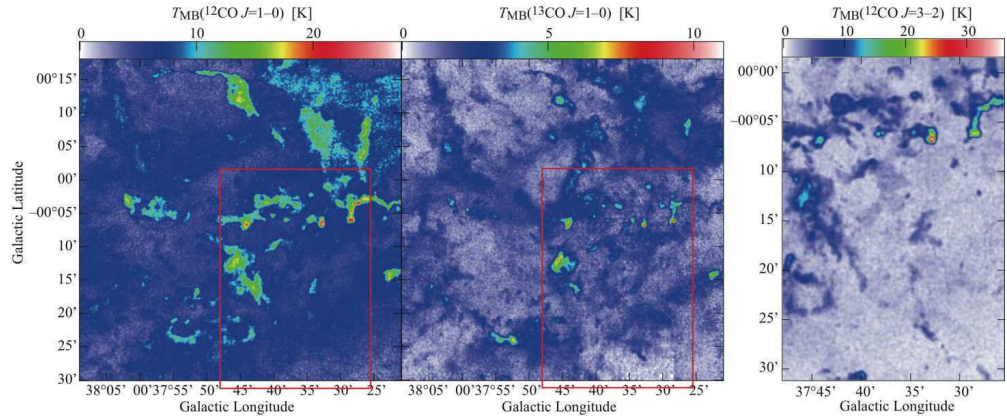}
\caption{Peak intensity ($T_{\rm MB}$) maps of the
  $^{12}{\rm CO}$ $J=1\mbox{--}0$, $^{13}{\rm CO}$ $J=1\mbox{--}0$, and
  $^{12}{\rm CO}$ $J=3\mbox{--}2$ lines.
  The region in which the $^{12}{\rm CO}$ $J=3\mbox{--}2$ line was
  observed is indicated by red rectangles in the $J=1\mbox{--}0$ maps.
  \label{fig:peaktemp}}
\end{figure}

The $^{12}{\rm CO}\;J=3\mbox{--}2$ data (Sawada et al.\ in prep.) are
compared with the $J=1\mbox{--}0$ lines.
The observations of the $J=3\mbox{--}2$ line were made with the
Atacama Sub\-millimeter Telescope Experiment (ASTE) 10-m telescope
at Pampa la Bola, Chile \citep{ezawa2004,kohno2005} in September 2005.
The region of $37\fdg 43 \lesssim l \lesssim 37\fdg 80$,
$-0\fdg 52 \lesssim b \lesssim +0\fdg 02$ (a part of the 45-m field of
view: see Fig.\ \ref{fig:peaktemp}) was mapped using the OTF technique.
The HPBW and $\eta_{\rm MB}$ of the telescope were $22\arcsec$ and 0.6,
respectively.
The data were reduced with {\it NOSTAR} and the reduced data cube has
an effective resolution of
$24\arcsec \times 24\arcsec \times 0.87\;{\rm km\;s^{-1}}$.
Details will be described in a forthcoming paper.
The peak intensity maps of the three lines are presented in
Fig.\ \ref{fig:peaktemp}.


\section{Results and Discussion}\label{sec:results}


\subsection{Velocity Channel Maps}\label{subsec:chmap}

Figs.\ \ref{fig:chmap12_1} and \ref{fig:chmap13_1} show the velocity
channel maps of the $^{12}{\rm CO}$ and $^{13}{\rm CO}$ lines,
respectively, which cover the velocity range
$v_{\rm LSR}=10\mbox{--}100\;{\rm km\;s^{-1}}$ with an interval of
5 ${\rm km\;s^{-1}}$.

The distribution of the emission in the channel maps significantly
changes with the velocity.
At the lowest velocity (10--25 ${\rm km\;s^{-1}}$) in
Fig.\ \ref{fig:chmap12_1}, low brightness $^{12}{\rm CO}$ emission
originating in the solar neighborhood is widespread in the field of
view.
Beyond 25 ${\rm km\;s^{-1}}$, the widespread emission vanishes.
Instead, bright ($T_{\rm MB} \gtrsim 10\;{\rm K}$) and spatially
confined structures begin to dominate at $\gtrsim 35\;{\rm km\;s^{-1}}$.
These structures have sharp boundaries and form clumps
(e.g., at $l\simeq 37\fdg 5$, $b\simeq -0\fdg 1$,
$v_{\rm LSR}=52.5\;{\rm km\;s^{-1}}$) and filaments (e.g., at the top of
the panel at 47.5 ${\rm km\;s^{-1}}$).
Beyond $\gtrsim 60\;{\rm km\;s^{-1}}$, the bright structures become less
prominent.
Though they remain up to 62.5 ${\rm km\;s^{-1}}$ (at the bottom of the
panel), low brightness extended emission starts to spread over the
field.
At the tangent velocity (75--90 ${\rm km\;s^{-1}}$), the low brightness
($\simeq 4\;{\rm K}$) emission almost fills the field of view,
except a lump of $\simeq 10\;{\rm K}$ emission at the top-right corner
of the field.
The high surface filling factor at the tangent velocity
can be attributed partly to the velocity crowding effect;
nevertheless, the lack of bright structures is a distinct difference 
from the velocity range of 40--60 ${\rm km\;s^{-1}}$
(Section \ref{subsec:bdf}).

The $^{13}{\rm CO}$ channel maps (Fig.\ \ref{fig:chmap13_1}) show
similar characteristics as those seen in the $^{12}{\rm CO}$ maps:
the bright ($\gtrsim 4\;{\rm K}$), compact structures dominate
at 40--65 ${\rm km\;s^{-1}}$, whereas the emission of 1--2 K is
widespread at 75--90 ${\rm km\;s^{-1}}$.
The $^{13}{\rm CO}$ maps have, in general, higher brightness contrast
than those of the $^{12}{\rm CO}$ line,
likely due to the lower optical depth.
One of the most extreme cases is the 62.5-${\rm km\;s^{-1}}$ channel.
The bright ($\gtrsim 4\;{\rm K}$) clumps seen in the $^{13}{\rm CO}$ map
are not very prominent in $^{12}{\rm CO}$,
implying the gas in the foreground with low excitation temperature
obscures the bright $^{12}{\rm CO}$ line of the clumps.

\begin{figure}
\epsscale{0.8}
\plotone{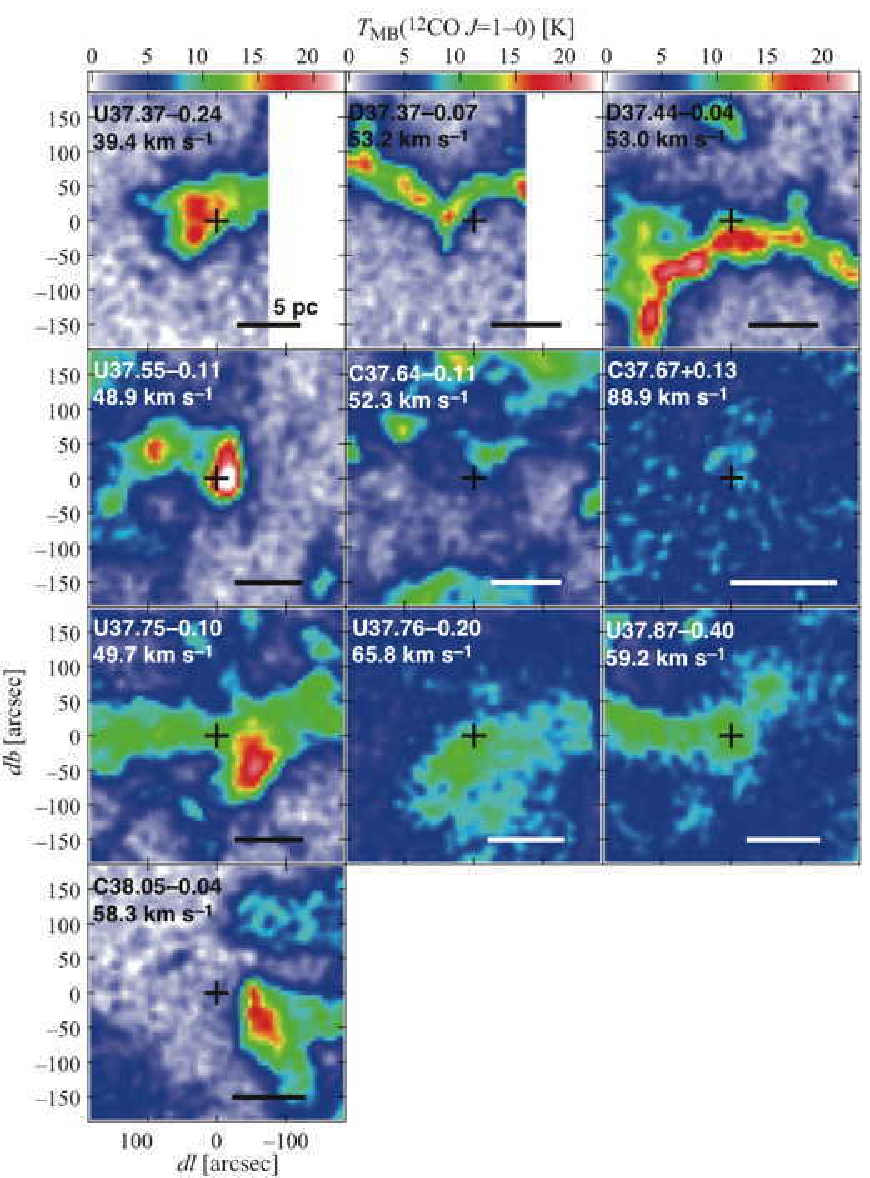}
\caption{The $^{12}{\rm CO}\;J=1\mbox{--}0$ peak intensity
  ($T_{\rm MB}$) maps around the \ion{H}{2} regions
  (shown as crosses) in \citet{anderson2009a}.
  The velocity range is $\pm 5\;{\rm km\;s^{-1}}$
  from the radial velocity of the \ion{H}{2} region,
  which is shown in each panel.
  The horizontal bars indicate the linear scale of 5 pc.
  Note that the coordinates in the catalog are given to
  the second decimal place in degrees:
  the sizes of the crosses are $\pm 0\fdg 005$ representing the
  uncertainty of the position.
  \label{fig:hii12}}
\end{figure}

\begin{figure}
\epsscale{0.8}
\plotone{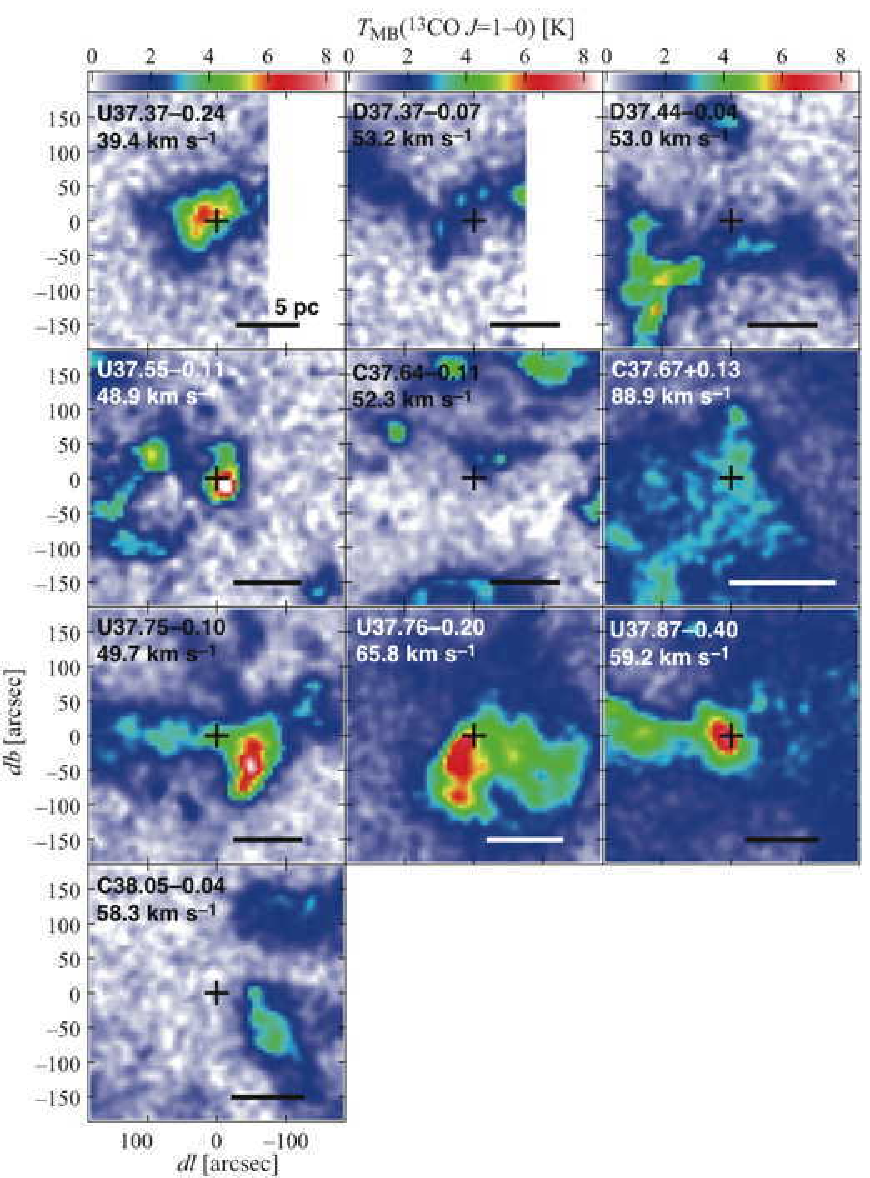}
\caption{Same as Fig.\ \ref{fig:hii12}, but for $^{13}{\rm CO}$.
  \label{fig:hii13}}
\end{figure}

\citet{anderson2009a} compiled catalogues of \ion{H}{2} regions
in the first Galactic quadrant, and determined the distances to them,
resolving the near-far distance ambiguity using
the \ion{H}{1} emission/absorption method and
the \ion{H}{1} self-absorption method.
Their catalog contains 10 \ion{H}{2} regions situated in our field of
view (Table \ref{tab:hii}).
Figs.\ \ref{fig:hii12} and \ref{fig:hii13} show the peak brightness maps
around the \ion{H}{2} regions in $^{12}{\rm CO}$ and $^{13}{\rm CO}$,
respectively.
Nine of them are associated with the bright ($>10\;{\rm K}$ in
$^{12}{\rm CO}$ and/or $>4\;{\rm K}$ in $^{13}{\rm CO}$) clumps or
filaments within a few pc.
In particular, 5 ultracompact \ion{H}{2} regions (whose names begin
with `U') are all tightly associated with the molecular clumps.
This is consistent with \citet{anderson2009b}.
One exception is C37.67$+$0.13 at the tangent velocity.
The $^{12}{\rm CO}$ around this object appears weak and extended,
probably due to the self-absorption.
The $^{13}{\rm CO}$ line is rather faint ($\approx 4\;{\rm K}$)
in comparison with the other regions.

Our sufficient ($\lesssim 1\;{\rm pc}$) spatial resolution reveals
that the spatial structure of molecular gas varies with the
radial velocity and thus with respect to the Galactic structure.
We classify the structures into the following two components,
the {\it \stateone} and the {\it \statetwo}.
The {\stateone} is bright ($T_{\rm MB}>10\;{\rm K}$ in $^{12}{\rm CO}$;
$>4\;{\rm K}$ in $^{13}{\rm CO}$) and spatially confined emission.
It appears as clumps and filaments in the channel maps, whose typical
size and mass are $1\arcmin$--$3\arcmin$ (3--8 pc at 9 kpc) and
$10^3$--$10^4 M_\sun$, respectively.
The {\statetwo} is diffuse (i.e., spatially extended) and fainter
($T_{\rm MB}\simeq 4\;{\rm K}$ in $^{12}{\rm CO}$
and $\simeq 1\;{\rm K}$ in $^{13}{\rm CO}$) emission.
The former is seen in the velocity range of
$\simeq 40\mbox{--}60 \;{\rm km\;s^{-1}}$, while the latter dominates
the solar neighborhood (10--25 ${\rm km\;s^{-1}}$) and at the tangent
velocity (75--90 ${\rm km\;s^{-1}}$).
The emission in the range 60--65 ${\rm km\;s^{-1}}$ is in-between
--- possibly a transition between or the mixture of them.

A number of studies have been performed in order to investigate the
distribution and physical conditions of the gas in the Milky Way Galaxy.
Most of them started from the decomposition of emission into individual,
discrete molecular clouds/clumps and determined their properties
(see Section \ref{sec:introduction}).
For example, \citet{scoville1987} identified 1427 {\it clouds} and 255
{\it hot cloud cores} from the Massachusetts-Stony Brook Survey data.
Among them 19 and 1 are in our field of view, respectively.
Their only hot cloud core is at
$(l, b, v)=(37\fdg 55, -0\fdg 10, 53\;{\rm km\;s^{-1}})$
and is a prototype of the {\stateone} in our classification.
Our higher-resolution and Nyquist-sampled maps reveal
a number of comparable structures (e.g., Fig.\ \ref{fig:peaktemp}).
We also find a significant extended molecular emission.
The previous studies might have missed
a considerable amount of emission because of the cloud identification,
i.e., the assumption that the molecular gas forms discrete objects.
The emission below the cloud boundary threshold can be
inevitably excluded from such analyses.
The {\statetwo} has a typical intensity ($\simeq 4\;{\rm K}$)
comparable to the boundary used by \citet{scoville1987} and
\citet{solomon1987}.
Thus a significant fraction of emission would have been overlooked in
their work.
In our data, 62\% and 48\% of the $^{12}{\rm CO}$ emission is below 4 K
in the velocity ranges of 40--60 and 75--90 ${\rm km\;s^{-1}}$,
respectively.
Though the {\stateone} emission will easily be identified as
clouds/clumps, it only accounts for a small fraction of the total gas
mass (Section \ref{subsec:bdi}).
In the following, we address the characteristics of the gas in a
quantitative fashion by using the {\it crude} observed line brightness,
rather than extracting clouds.


\subsection{Brightness Distribution Function}\label{subsec:bdf}

\begin{figure}
\epsscale{0.8}
\plotone{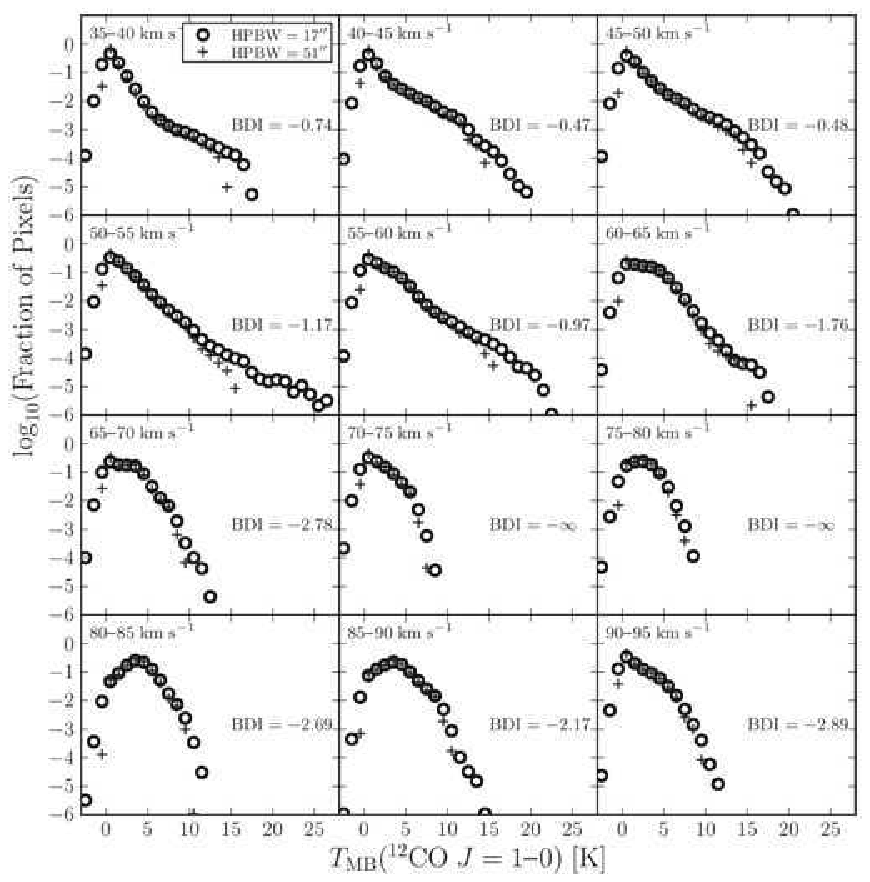}
\caption{Histogram of the $^{12}{\rm CO}$ brightness temperature.
  The horizontal axis is the brightness temperature $T_{\rm MB}$ [K],
  and the vertical axis is the fraction of pixels in each 1-K brightness
  bin.
  Open circles and crosses represent the original resolution
  ($\mbox{effective HPBW}=17\arcsec$) and 3-times smoothed ($51\arcsec$)
  data, respectively.
  \label{fig:bdf_12}}
\end{figure}

\begin{figure}
\epsscale{0.8}
\plotone{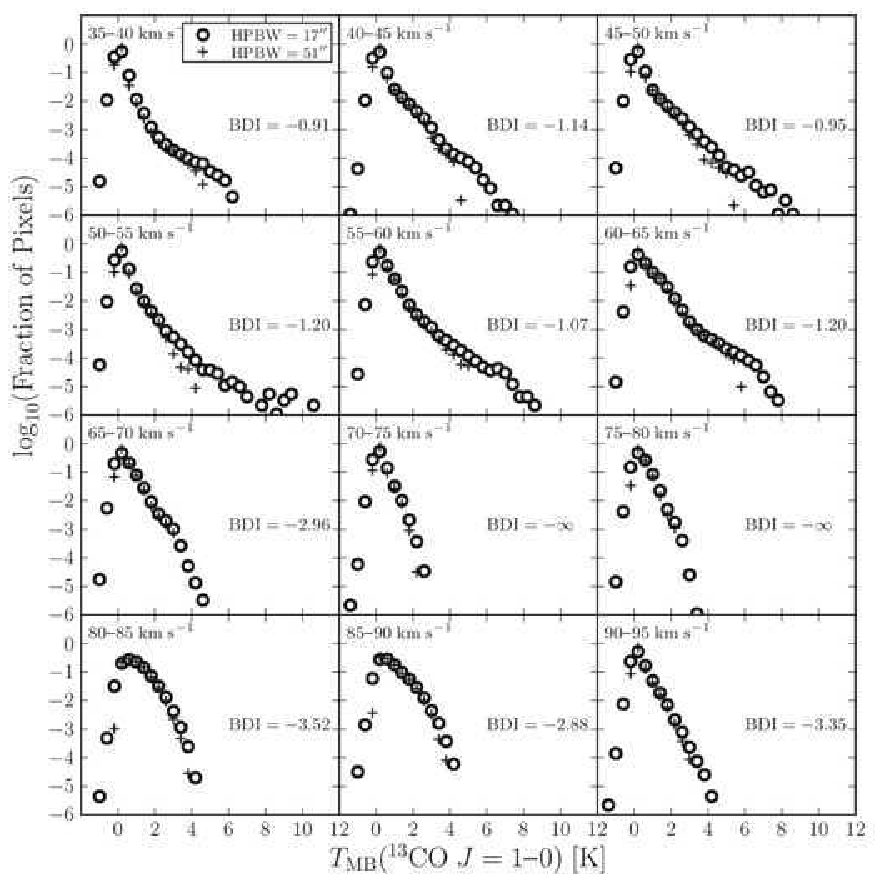}
\caption{Same as Fig.\ \ref{fig:bdf_13}, but for $^{13}{\rm CO}$.
  The width of each brightness bin is 0.4 K.
  \label{fig:bdf_13}}
\end{figure}

There is clear variation of spatial structure of the gas
with the radial velocity.
Here we use histograms of the brightness temperatures,
or the {\it Brightness Distribution Functions} (BDFs),
which visualize the spatial structure of the gas without separating
structures into arbitrary clouds.
Figs.\ \ref{fig:bdf_12} and \ref{fig:bdf_13} present the BDFs
of $^{12}{\rm CO}$ and $^{13}{\rm CO}$, respectively.
The data with the
$6\arcsec \times 6\arcsec \times 1.25\;{\rm km\;s^{-1}}$
grid are divided into 5-${\rm km\;s^{-1}}$ velocity channels,
and the number of $l$-$b$-$v$ pixels in each 1 K ($^{12}{\rm CO}$) and
0.4 K ($^{13}{\rm CO}$) brightness bin is plotted
(normalized by the total number of pixels in each velocity channel).
In each panel the {\it Brightness Distribution Index} is also shown,
which we introduce in the following subsection in order to quantify
the characteristics of BDF.

The BDF clearly reflects the amount of the two components
of molecular gas described above.
The $^{12}{\rm CO}$ BDF in the
$v_{\rm LSR} = 35\mbox{--}40\;{\rm km\;s^{-1}}$
channel (Fig.\ \ref{fig:bdf_12}) shows a sharp peak in the 0--1 K
brightness bin.
This corresponds to the fact that a large portion of the field of view
is almost emission-free (Fig.\ \ref{fig:chmap12_1}).
The high-brightness tail
represents the {\stateone} in the right-hand side
of the field (the most prominent structure is at
$l=37\degr 23\arcmin$, $b=-0\degr 14\arcmin$).
As the velocity goes up to $v_{\rm LSR}\approx 55\;{\rm km\;s^{-1}}$,
the high-brightness tail is even more populated, and
the peak brightness also increases.
It corresponds to the {\stateone} structures in the velocity channel
maps, with increased numbers (i.e., surface filling factors) and peak
brightness.
The peak of the BDF at $\simeq 0\;{\rm K}$ is still prominent,
reflecting the relatively small surface filling factor of the CO
emission.
At $\approx 60\;{\rm km\;s^{-1}}$, the 0-K peak starts to drop,
and another remarkable component, the shoulder at $\approx 4\;{\rm K}$,
emerges.
It reflects the {\statetwo} which fills the bottom half of the field.
At $\approx 75\;{\rm km\;s^{-1}}$,
the high-brightness tail truncates,
while the 4-K shoulder still exists.
At the tangent velocity (75--90 ${\rm km\;s^{-1}}$), the 0-K peak is
no longer prominent, and the 4-K shoulder turns into a peak.
This transition is obvious in the channel maps, in which {\statetwo}
fills almost the whole field of view.
Beyond 95 ${\rm km\;s^{-1}}$, the 0-K peak reappears
since the surface filling factor of the {\statetwo} decreases.
The BDF, with a 4-K shoulder and truncation toward high brightness,
is similar to that at 65--75 ${\rm km\;s^{-1}}$.

The $^{13}{\rm CO}$ BDF shows a similar trend --- high-brightness
($\gtrsim 4\;{\rm K}$) tail at 40--65 ${\rm km\;s^{-1}}$,
truncation beyond 65 ${\rm km\;s^{-1}}$,
and a shoulder at $\approx 1\;{\rm K}$ at the tangent velocity.
The $^{13}{\rm CO}$ BDF has, in general, a more prominent 0-K peak
and sharper truncation toward the high-brightness tail,
which come from the higher brightness contrast
in the $^{13}{\rm CO}$ channel maps (Section \ref{subsec:chmap}).

We suggest that the difference of the BDF between the Sgr arm
($40\mbox{--}60 \;{\rm km\;s^{-1}}$) and the inter-arm region
($75\mbox{--}90 \;{\rm km\;s^{-1}}$) is due to a change of
gas properties.
There are, however, a couple of potential issues that we should
consider, which might cause apparent variation between the velocity
components:
(a) distance from us (i.e., linear resolution), and
(b) velocity crowding.

In order to test (a), we convolve the maps with a Gaussian to degrade
the resolution by a factor of 3
(equivalent to ${\rm HPBW}=51\arcsec$)
and see how the spatial resolution affects the map's appearance and
thus the BDF.
The BDFs from the convolved maps
(crosses in Figs.\ \ref{fig:bdf_12} and \ref{fig:bdf_13})
confirm the features discussed above.
This implies that a near-far distance ambiguity up to a factor of
$\approx 3$ (Section \ref{sec:field}) does not affect the result.

As for (b), the line-of-sight path length in a 10-${\rm km\;s^{-1}}$
bin for the tangent component is $\approx 2.5$ times larger than those
for the near- and far-side Sgr arm (Section \ref{sec:field}).
This increases the $(l, b, v)$-volume filling factor of the emitting
region in the tangent component,
which may have turned the 4-K shoulder in the $^{12}{\rm CO}$ BDF into
the peak seen at the tangent velocity.
The absence of bright ($\gtrsim 10\;{\rm K}$) $^{12}{\rm CO}$ emission
may possibly be attributed to velocity crowding and self-absorption.
However, the optically thin $^{13}{\rm CO}$ emission
also lacks bright ($\gtrsim 4\;{\rm K}$) structures at this velocity.
This supports that the BDF in the inter-arm region differs intrinsically
from that in the Sgr arm.


\subsection{Brightness Distribution Index}\label{subsec:bdi}

The BDFs clearly characterize the variation of
the spatial structure, showing the bright, compact clumps/filaments
and more extended component of molecular gas.
We introduce the {\it brightness distribution index} (BDI) to
characterize the BDF with one number,
which can be correlated with other parameters, such as
the physical conditions of the gas
and star-forming activity.
The BDI is defined as the flux ratio
of the bright emission to the low-brightness emission:
\begin{eqnarray}
{\mit BDI} &=& \log_{10} \left(
  \frac{\int_{T_2}^{T_3} T\cdot B(T) dT}{\int_{T_0}^{T_1} T\cdot B(T) dT}
 \right) \nonumber\\
 &=& \log_{10} \left(
  \frac{\sum_{T_2<T[i]<T_3}T[i]}{\sum_{T_0<T[i]<T_1}T[i]}
 \right),
 \label{eq:bdi}
\end{eqnarray}
where the BDF is denoted as $B(T)$; $T_0$, $T_1$, $T_2$, and $T_3$ are
the brightness thresholds; and $T[i]$ is the brightness of the $i$-th
pixel.
We note that the flux is roughly proportional to the mass.

In this paper we adopt $(T_0, T_1, T_2, T_3) = (3, 5, 10, \infty)$ for
$^{12}{\rm CO}$ and $(1, 1.5, 4, \infty)$ for $^{13}{\rm CO}$.
These thresholds are somewhat arbitrary, but are empirically chosen
so that the numerator and
the denominator of Eq.\ (\ref{eq:bdi}) represent the gas of the
typical brightness of the B and D components that are evident in
the channel maps (Section \ref{subsec:bdf}), respectively.
The typical molecular gas temperature that is determined in a wider
Galactic plane $^{12}{\rm CO}$ survey is $\simeq 10$ K
\citep{scoville1987in};
subtracting the cosmic microwave background temperature,
the average brightness temperature would be $\simeq 7$ K.
Our choice of the thresholds represents the brightness appreciably
below and above this average over the large area in the Galactic plane.
The thresholds for $^{13}{\rm CO}$ are adjusted correspondingly
to pick out the similar regions in the maps (Figs.\ \ref{fig:chmap12_1}
and \ref{fig:chmap13_1}).
In the following we demonstrate the utility of the BDI
to characterize the spatial structure, despite that the thresholds
can be specific to the line of sight we observed.
An important future work is to revisit the choice of
these parameters when more examples become available.

The BDIs for $^{12}{\rm CO}$ and $^{13}{\rm CO}$
within the 5-${\rm km\;s^{-1}}$ velocity bins
are shown in Figs.\ \ref{fig:bdf_12} and \ref{fig:bdf_13}, respectively.
At $v_{\rm LSR} \simeq 40\mbox{--}60 \;{\rm km\;s^{-1}}$
the BDIs are higher ($-1.2$--$-0.5$ in $^{12}{\rm CO}$,
$-1.2$--$-0.9$ in $^{13}{\rm CO}$),
while at the $80\mbox{--}90 \;{\rm km\;s^{-1}}$ BDIs are lower
($-2.7$--$-2.2$ in $^{12}{\rm CO}$, $-3.5$--$-2.9$ in $^{13}{\rm CO}$).
The fraction of bright emission varies with velocity.
It is also remarkable that
even at the velocities with a high BDI (i.e., in spiral arm),
only a small fraction of the gas is composed in the {\stateone}.
That is, despite the highest $^{12}{\rm CO}$ BDI ($-0.47$)
at 45--50 ${\rm km\;s^{-1}}$,
the flux of the $T_{\rm MB} > T_2$ emission amounts to only
one third of the $T_0 < T_{\rm MB} < T_1$ emission, or,
6.6\% of the total flux.
The mass fraction of the {\stateone} gas is even lower
at other radial velocities.

\begin{figure}
\epsscale{0.8}
\plotone{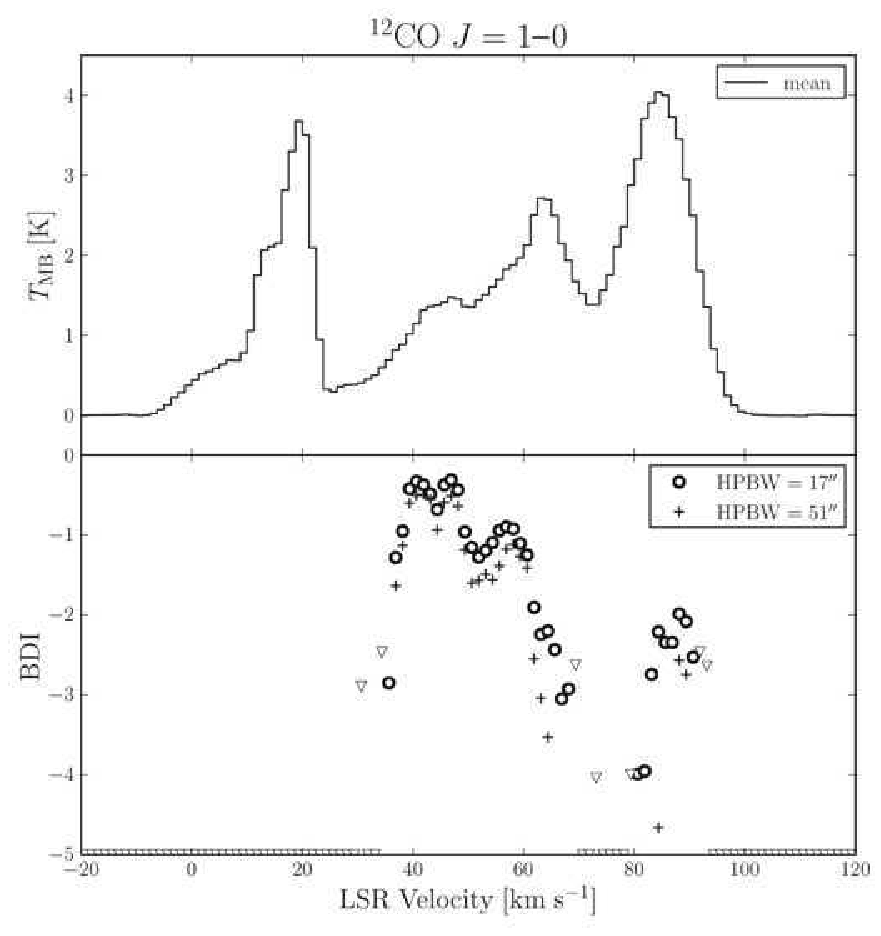}
\caption{(Top) The mean brightness of the $^{12}{\rm CO}$ line.
  (Bottom) The velocity profile of the BDI.
  Open circles and crosses represent the original resolution and
  3-times smoothed data, respectively.
  Open triangles are the upper limit of the BDI, defined as
  $\log_{10} \{
    {\sum_{T_2<(T[i]+3\sigma)}(T[i]+3\sigma)}/{\sum_{T_0<T[i]<T_1}T[i]}
  \}$ ($\sigma$ is the RMS noise of the map),
  for the original resolution data.
  \label{fig:bdi_prof_12}}
\end{figure}

\begin{figure}
\epsscale{0.8}
\plotone{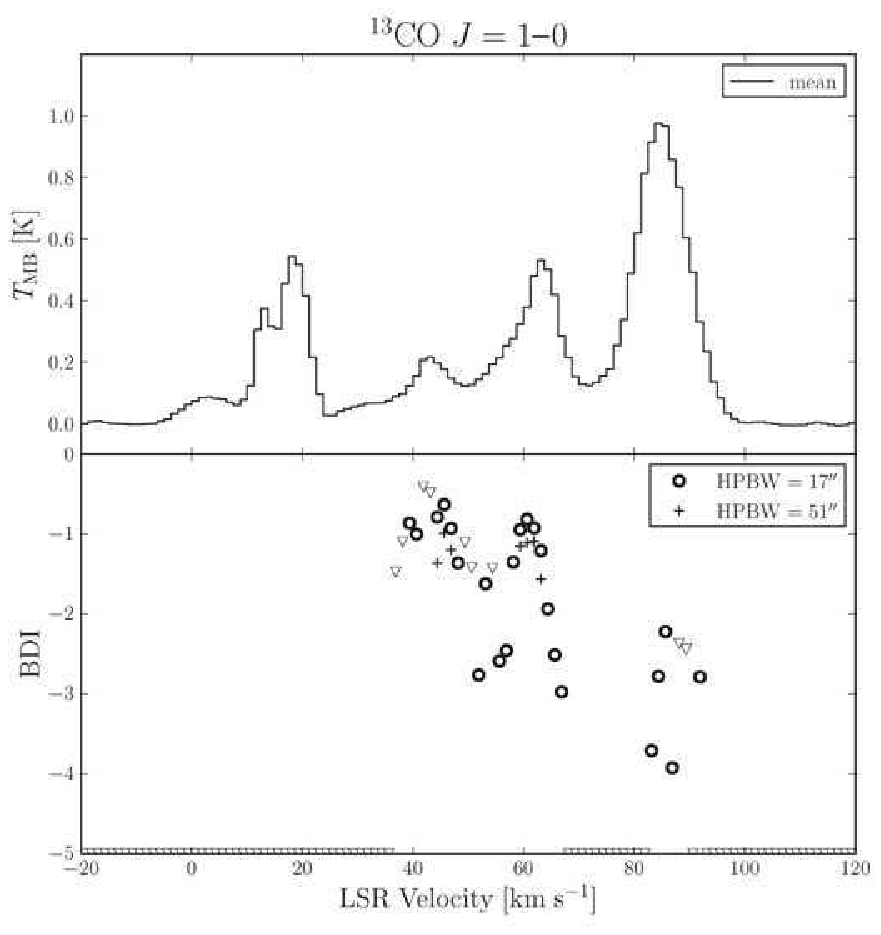}
\caption{Same as Fig.\ \ref{fig:bdi_prof_12}, but for $^{13}{\rm CO}$.
  The region observed in Period 1 is excluded since the velocity
  coverage is narrow.
  \label{fig:bdi_prof_13}}
\end{figure}

The variation of BDI as a function of radial velocity (i.e.,
velocity profiles of BDIs) for $^{12}{\rm CO}$ and $^{13}{\rm CO}$
are presented in Figs.\ \ref{fig:bdi_prof_12} and \ref{fig:bdi_prof_13},
along with the mean brightness in the field.
The line brightness shows 4 prominent peaks at
$20\;{\rm km\;s^{-1}}$ (solar neighborhood),
$45\;{\rm km\;s^{-1}}$ (near-side Sgr arm),
$65\;{\rm km\;s^{-1}}$ (far-side Sgr arm), and
$85\;{\rm km\;s^{-1}}$ (tangent velocity).
The BDI in $^{12}{\rm CO}$ is high in the velocity range
40--60 ${\rm km\;s^{-1}}$ (i.e., Sgr arm),
with a steep decrease beyond 60 ${\rm km\;s^{-1}}$.
The BDI is low at 80--90 ${\rm km\;s^{-1}}$.
In other velocity ranges the BDI is infinitely small because of the lack
of $\ge T_2$ emission.
The $^{13}{\rm CO}$ BDI behaves similarly to that of $^{12}{\rm CO}$:
it is high in 40--60 ${\rm km\;s^{-1}}$ with peaks at 45 and 60
${\rm km\;s^{-1}}$ (the Sgr arm),
and is low in 80--90 ${\rm km\;s^{-1}}$ (the inter-arm region).
Molecular gas is extended with little brightness variation
in the inter-arm region, and bright, compact structures emerge
in spiral arms.


\subsection{Comparison with CO Intensity Ratios}\label{subsec:ratio}

The $^{12}{\rm CO}\;J=3\mbox{--}2$ data are available for a smaller
field (Fig.\ \ref{fig:peaktemp}), $\approx 1/4$ of the 45-m field
of view.
The BDI, the $^{12}{\rm CO}\;J=1\mbox{--}0$ brightness, the
$T_{\rm MB}(^{12}{\rm CO}\;J=3\mbox{--}2)/
T_{\rm MB}(^{12}{\rm CO}\;J=1\mbox{--}0)$
ratio [$R_{3\mbox{--}2/1\mbox{--}0}(^{12}{\rm CO})$], and the
$T_{\rm MB}(^{13}{\rm CO}\;J=1\mbox{--}0)/
T_{\rm MB}(^{12}{\rm CO}\;J=1\mbox{--}0)$
ratio [$R_{13/12}(J=1\mbox{--}0)$]
are shown in Fig.\ \ref{fig:bdi_prof_12_asteregion}.
The overall characteristics of the $^{12}{\rm CO}$ brightness and the
BDI are similar to those in the whole 45-m field of view
(Fig.\ \ref{fig:bdi_prof_12}).
The $R_{3\mbox{--}2/1\mbox{--}0}(^{12}{\rm CO})$ is the highest at the
radial velocity of 40--45 ${\rm km\;s^{-1}}$ and tends to decrease
toward the higher velocity.
It has local peaks at 40--45, 55, and 85 ${\rm km\;s^{-1}}$.
The first two peaks agree with those of the BDI.
If we assume that the brightness peaks at $\approx 47$ and 63
${\rm km\;s^{-1}}$ trace the near- and far-side Sgr arm respectively,
these BDI and $R_{3\mbox{--}2/1\mbox{--}0}(^{12}{\rm CO})$ peaks
are both offset from the
corresponding brightness peaks toward the lower velocity
by several ${\rm km\;s^{-1}}$ (discussed in Section \ref{subsec:arm}).
On the other hand, the radial velocity of the third peak coincides with
that of the brightness peak.
The $R_{13/12}(J=1\mbox{--}0)$ generally correlates with the brightness,
and traces the optical depth (the column density) of the gas.
Thus the $^{12}{\rm CO}$
brightness peaks correspond to the peaks of molecular gas distribution
along the radial velocity.

The sets of the ratios
$(R_{3\mbox{--}2/1\mbox{--}0}(^{12}{\rm CO}), R_{13/12}(J=1\mbox{--}0))$
are calculated for the two components
(1) $v_{\rm LSR} = 40\mbox{--}60\;{\rm km\;s^{-1}}$,
$T_{\rm MB}(^{12}{\rm CO}) > 10\;{\rm K}$ and
(2) $v_{\rm LSR} = 70\mbox{--}80\;{\rm km\;s^{-1}}$\footnote{A slightly
lower velocity than the tangent velocity is chosen, since the emission
in the tangent is heavily affected by the velocity crowding.},
$T_{\rm MB}(^{12}{\rm CO}) = 3\mbox{--}5\;{\rm K}$.
They represent the B and D components.
The ratios are derived as $(\ge 0.63\pm 0.03, 0.23\pm 0.02)$ and
$(0.40\pm 0.11, 0.16\pm 0.05)$\footnote{We took account of the fact that
the brightness temperature of spatially extended structures
is overestimated in $T_{\rm MB}$ scale
(Appendix \ref{app:efficiency}).}, respectively.
The errors quoted are estimated from
baseline uncertainties (i.e., constant offset of $1\sigma$ is
assumed in all spectra over all channels, which we consider as a
very conservative estimate) and the random noise.

\begin{figure}
\epsscale{0.8}
\plotone{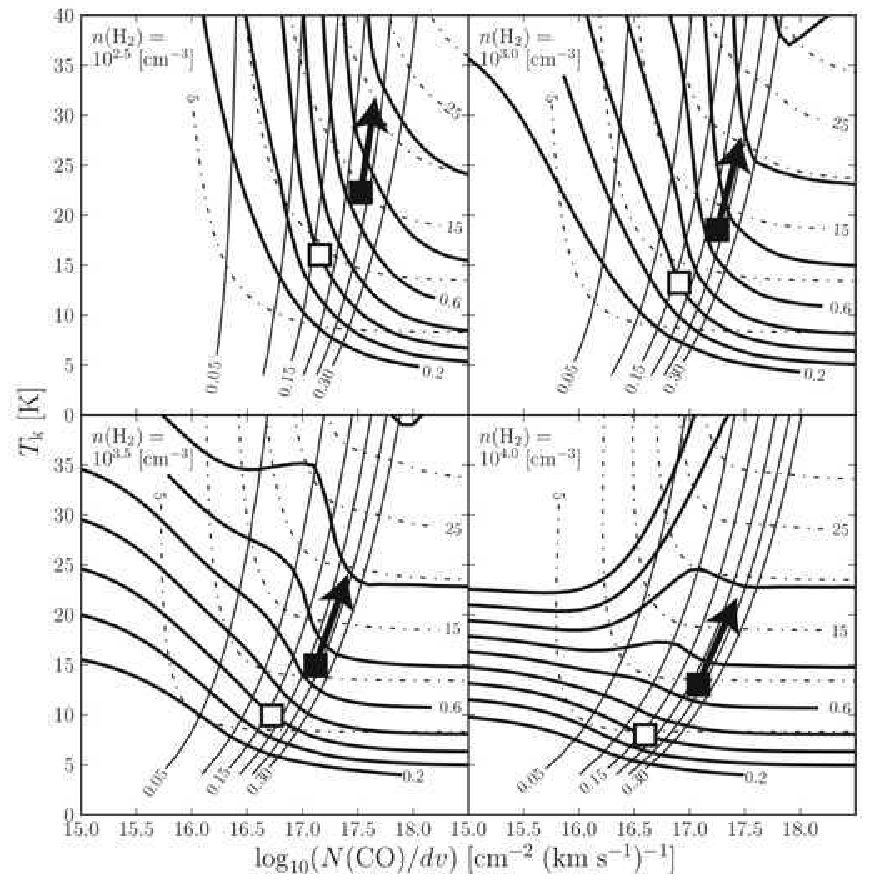}
\caption{The LVG results for $n({\rm H_2}) = 10^{2.5}$, $10^{3.0}$,
  $10^{3.5}$, and $10^{4.0}\; {\rm cm^{-3}}$.
  The $R_{3\mbox{--}2/1\mbox{--}0}(^{12}{\rm CO})$,
  $R_{13/12}(J=1\mbox{--}0)$, and the $^{12}{\rm CO}\;J=1\mbox{--}0$
  brightness temperature are drawn as thick solid lines (the contour
  levels are 0.2, 0.3, 0.4, ...), thin solid lines
  (0.05, 0.10, 0.15, ...), and dashed lines (5, 10, 15, ... [K]),
  respectively.  Estimated physical conditions for the two components
  ({\stateone} and {\statetwo}) are marked as the filled and
  open squares.
  \label{fig:lvg}}
\end{figure}

In order to estimate qualitatively what the main difference
between the two component is, we compare the obtained intensity
ratios with simple model calculations.
Here we use the large-velocity-gradient
\citep[LVG;][]{goldreich1974,scoville1974} model and
adopt the one-zone assumption: i.e., the emission lines originate
in a homogeneous volume of the gas.
More detailed analyses of the physical conditions of the gas
by using more lines will be reported in a forthcoming paper.
Fig.\ \ref{fig:lvg} shows the result of LVG calculations
made with {\it RADEX} \citep{vandertak2007}.
The $R_{3\mbox{--}2/1\mbox{--}0}(^{12}{\rm CO})$,
$R_{13/12}(J=1\mbox{--}0)$, and the $^{12}{\rm CO}\;J=1\mbox{--}0$
brightness temperature were calculated as functions of the kinetic
temperature of the gas ($T_{\rm k}$) and the column density of
$^{12}{\rm CO}$ molecules per unit velocity width [$N({\rm CO})/dv$].
The number density of molecular hydrogen [$n({\rm H_2})$] was assumed to
be $10^{2.5}$, $10^{3.0}$, $10^{3.5}$, and $10^{4.0}$ ${\rm cm^{-3}}$.
The $^{12}{\rm C}/^{13}{\rm C}$ abundance ratio at a Galacto\-centric
distance of 6 kpc was estimated to be 40--55
\citep{langer1993,savage2002,milam2005}.
Here we adopted 50 for the calculation.
The physical conditions which reproduce the observed intensity ratios
are plotted as the filled and open squares.
The $T_{\rm k}$ of the two components are estimated to be
$\ge 13\mbox{--}22$ and 8--16 K if the $n({\rm H_2})$ is in the range of
$10^{2.5}\mbox{--}10^{4.0}\;{\rm cm^{-3}}$
(note that $R_{3\mbox{--}2/1\mbox{--}0}(^{12}{\rm CO})$, and thus
the derived $T_{\rm k}$, of the {\stateone} gas is a lower limit).
The gas in the {\stateone} is found to be warmer than
that of the {\statetwo}, as expected.
Although no tight constraints on the gas density can be given,
the possibility of the high ($n({\rm H_2}) \gtrsim 10^4\;{\rm cm^{-3}}$)
density of the {\statetwo} is rejected
because the observed brightness cannot be reproduced even if the
surface filling factor is unity.


\subsection{Comparison with \ion{H}{2} Regions}

\begin{figure}
\epsscale{0.8}
\plotone{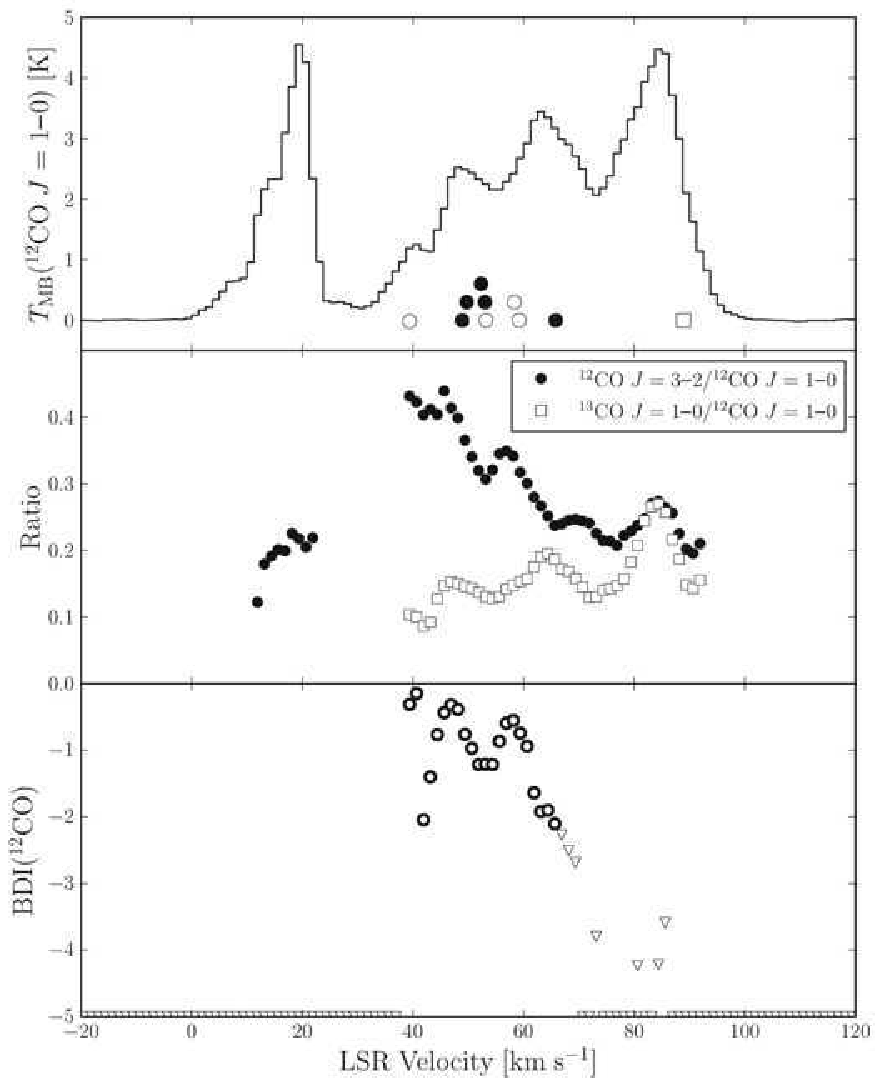}
\caption{(Top) The line profile of $^{12}{\rm CO}\;J=1\mbox{--}0$.
  The radial velocities of the \ion{H}{2} regions taken from
  \citet{anderson2009a} are also shown as
  circles (at the far side), and a square (at the tangent point).
  The \ion{H}{2} regions inside the ASTE field of view are drawn as
  filled symbols, while the others are open.
  (Middle)
  The $^{12}{\rm CO}\;J=3\mbox{--}2/^{12}{\rm CO}\;J=1\mbox{--}0$ and
  $^{13}{\rm CO}\;J=1\mbox{--}0/^{12}{\rm CO}\;J=1\mbox{--}0$ intensity
  ratios.
  Note that the
  $^{12}{\rm CO}\;J=3\mbox{--}2/^{12}{\rm CO}\;J=1\mbox{--}0$ ratio
  is most probably underestimated (see text).
  (Bottom) The BDI in $^{12}{\rm CO}\;J=1\mbox{--}0$,
  inside the region $37\fdg 43 \lesssim l \lesssim 37\fdg 80$,
  $-0\fdg 50 \lesssim b \lesssim +0\fdg 02$.
  \label{fig:bdi_prof_12_asteregion}}
\end{figure}

The \ion{H}{2} region catalog compiled by \citet{anderson2009a} contains
10 samples in our field of view (Table \ref{tab:hii}: see
Section \ref{subsec:chmap}).
The radial velocities of these \ion{H}{2} regions are plotted in
Fig.\ \ref{fig:bdi_prof_12_asteregion}.
The ones in the ASTE field of view are shown in filled symbols,
while the others are open.
Eight \ion{H}{2} regions are located at the far side 
(circles in Fig.\ \ref{fig:bdi_prof_12_asteregion}, top), and they are
mostly concentrated at $\approx 50\mbox{--}60\;{\rm km\;s^{-1}}$.
This coincides with the velocity range where
the BDI and the $R_{3\mbox{--}2/1\mbox{--}0}(^{12}{\rm CO})$ are high.

These 8 \ion{H}{2} regions at the far side
($50\mbox{--}60\;{\rm km\;s^{-1}}$) are most likely associated with
the far-side Sgr arm (i.e., the peak of the brightness
at $\approx 63\;{\rm km\;s^{-1}}$).
Thus the \ion{H}{2} regions in our line of sight have radial velocities
systematically lower than that of the molecular spiral arm.
We note again that the radial velocities of the BDI and
$R_{3\mbox{--}2/1\mbox{--}0}(^{12}{\rm CO})$ peaks
are offset from those of the CO brightness maxima
toward the lower velocity (Section \ref{subsec:ratio}).
All BDI and $R_{3\mbox{--}2/1\mbox{--}0}(^{12}{\rm CO})$ peaks
and \ion{H}{2} regions are offset in velocity from the spiral arm.


\subsection{Molecular Gas and Spiral Arm}\label{subsec:arm}

We identified two distinct components of the molecular gas,
the {\stateone} and the {\statetwo},
based on high-resolution, wide-field mapping observations
of the Galactic plane.
The {\stateone} emission is prominent at the velocities
of the spiral arms,
although its flux (mass) fraction of the total flux is small.
The {\statetwo} (and even fainter emission) is
the majority of the molecular mass,
dominating the emission both at the spiral arm and
at the inter-arm velocities.
We have introduced the BDF and the BDI to quantify the difference,
and demonstrated that the BDI
characterizes the structural variation of the gas
along the line of sight: i.e., the BDI is high in the
spiral arm velocities and low outside.
The {\stateone} emission corresponds to the {\it ``warm''} or
{\it ``hot''} clouds defined in \citet{solomon1985} and
\citet{scoville1987}.
Our analysis confirms their results that the {\it ``warm''} or
{\it ``hot''} clouds are concentrated on the $l$-$v$ loci of
the spiral arms and resolves them as bright clumps and filaments.
The advantages of our study are the following:
The high-resolution observations revealed that the {\statetwo} emission
is faint and spatially extended, and therefore was missed
in the ``cloud identification'' scheme.
This component consists about half the mass in our field
and is quite substantial.
Our method is free from the process of cloud identification
and takes into account such a component as well as bright and clumpy
(the {\stateone}) emission.
Furthermore, the new parameter, BDI, characterizes the gas structure and
is directly comparable with other tracers of the spiral arms (e.g.,
\ion{H}{2} regions) and the physical conditions of the gas (e.g.,
line ratios).

Extended emission in CO $J=1\mbox{--}0$ has been found in external
galaxies, especially in their inter-arm regions
\citep[e.g.,][]{adler1992,koda2009}.
However, at their current resolutions even with interferometers
($> 100$ pc), it remains unclear if these components are an
ensemble of small unresolved giant molecular clouds or
truly-extended emission.
Our results are drawn at a very high spatial resolution of
$\simeq 0.5$ pc, and are therefore different from the extragalactic
results both qualitatively and quantitatively.
The relation between such small structures and kpc-scale
galactic structures that we find in Figs.\ \ref{fig:chmap12_1} and
\ref{fig:chmap13_1} is the new finding.
The Atacama Large Millimeter/submillimeter Array will bridge
the gap between these studies with its ability to produce
high-fidelity images of pc-sized structures over nearby galaxies.

The peaks of the BDI coincide in velocity with
the $R_{3\mbox{--}2/1\mbox{--}0}(^{12}{\rm CO})$ local maxima.
The {\stateone} gas, which makes the BDI high,
shows a high $R_{3\mbox{--}2/1\mbox{--}0}(^{12}{\rm CO})$ ratio
and is warm, as opposed to the {\statetwo} gas.
The BDI peaks also coincide with \ion{H}{2} regions,
but are offset toward lower velocities
from the maxima of the CO brightness (the near- and far-side Sgr arm).
These results indicate that the distribution of the high-BDI gas
and \ion{H}{2} regions is shifted from the molecular spiral arm
by several ${\rm km\;s^{-1}}$.

The offset in velocity between the high-BDI gas and molecular spiral arm
implies that the high-BDI gas is located in the outer (i.e., larger
Galacto\-centric radius: see Fig.\ \ref{fig:lvpunch}b) side of the
spiral arm.
If we assume the 220-${\rm km\;s^{-1}}$ flat rotation of the Galaxy,
the velocity offset (several ${\rm km\;s^{-1}}$) translates to
$\simeq 500\;{\rm pc}$ in space.
The corotation radius of spiral arms in the Milky Way Galaxy
is likely outside the solar circle\footnote{For example,
\citet{bissantz2003} studied the gas dynamics in the Galaxy
and concluded that the pattern speed of the spiral arms is
$\approx 20\;{\rm km\;s^{-1}\;kpc^{-1}}$.}.
Therefore the gas at a Galacto\-centric radius of 6 kpc revolves faster
than the spiral pattern, by $\approx 100\;{\rm km\;s^{-1}}$.
Hence, the gas with high BDI and high
$R_{3\mbox{--}2/1\mbox{--}0}(^{12}{\rm CO})$ and star-forming regions
are located on the downstream side of the spiral arm
(CO brightness peak) in our field of view.
If we assume a pitch angle of the Sgr arm of $12\degr$
\citep{georgelin1976}, the drift timescale for the offset is
$\sim 20\;{\rm Myr}$.
This is consistent with the timescale found in other galaxies
\citep{egusa2009}.
Note that the velocity offset may partially be attributed to
the effect of radiative transfer
(e.g., the {\stateone} gas in the spiral arm velocity with high
$R_{3\mbox{--}2/1\mbox{--}0}(^{12}{\rm CO})$ may be obscured
by the bulk {\statetwo} gas) and needs to be verified by
using optically thinner lines.
Nevertheless, the asymmetry of BDI, 
$R_{3\mbox{--}2/1\mbox{--}0}(^{12}{\rm CO})$, and
the distribution of \ion{H}{2} regions with respect to the
arm velocities indicates that the trend is real.

We treated the tangent-velocity gas
as a prototype of the gas in the inter-arm region.
There are, however, some caveats.
First, this line of sight shows slightly more emission at the tangent
velocity compared with the neighboring longitudes
\citep[e.g., Fig.\ 3 in][]{dame2001}.
This area is a molecular-rich inter-arm region.
Second, the BDI is higher compared with
the other velocity ranges corresponding to inter-arm regions:
i.e., between the far-side Sgr arm and the tangent velocity
(70--80 ${\rm km\;s^{-1}}$) and
the solar neighborhood ($\simeq 20\;{\rm km\;s^{-1}}$).
The emission at the tangent velocity has been proposed,
though not proven, as a special structure, a spur
that extends from a spiral arm into the inter-arm region
\citep[e.g.,][]{dame1986}.
\citet{sakamoto1997} found that there was no enhancement of the
$^{12}{\rm CO}\;J=2\mbox{--}1/^{12}{\rm CO}\;J=1\mbox{--}0$ ratio and
deduced a lower gas density than the average over a much larger region.
\citet{koda2009} performed high-resolution mapping observations
of the $^{12}{\rm CO}\;J=1\mbox{--}0$ line in the entire disk of M51
and detected spurs in the inter-arm regions.
The gas at our tangent velocity could be a relatively molecular-rich
portion of the inter-arm region.
Still, the clear difference of gas structure between spiral arms
and inter-arm regions is striking.
The richness of molecular gas is not the only determinant of
gas structure.
There is a correlation with large galactic structure.


\section{Conclusions}\label{sec:conclusions}

We performed mapping observations of a $0\fdg 8\times 0\fdg 8$ field
on the Galactic plane in the $^{12}{\rm CO}$ and $^{13}{\rm CO}$
$J=1\mbox{--}0$ lines.
The high-resolution maps resolve the spatial structure of the emission
down to $\lesssim 1\;{\rm pc}$ and clearly show its variation with the
radial velocity, and therefore, between spiral arm and inter-arm
regions.
The bright and spatially confined emission ({\stateone})
is prominent in the Sgr arm, while the fainter, diffuse emission
({\statetwo}) dominates in the inter-arm regions.
We investigated the characteristics of the gas
and revealed the followings:

\begin{itemize}
  \item 
    The typical size and mass of the {\stateone} structures
    are $1\arcmin\mbox{--}3\arcmin$ (3--8 pc at 9 kpc)
    and $10^3$--$10^4 M_\sun$, respectively;
    and it contains only a small fraction of the total gas mass.
    The {\stateone} exists predominantly in spiral arms.
    The {\statetwo} is widespread and dominant in the inter-arm region.
    It also exists in the arm as well.
  \item The BDFs characterize the variation of the spatial structure of
    the gas.
    We also defined a new parameter, BDI, as the flux
    ($\sim {\rm mass}$) ratio between the {\stateone} and {\statetwo},
    in order to characterize the spatial structure.
  \item High BDI coincides with
    high $R_{3\mbox{--}2/1\mbox{--}0}(^{12}{\rm CO})$.
    Thus the high BDI component contains warm/hot gas.
  \item High BDI also coincides with \ion{H}{2} regions.
    Almost all \ion{H}{2} regions are associated with the
    {\stateone} emission.
  \item The high-BDI gas (and thus, the high
    $R_{3\mbox{--}2/1\mbox{--}0}(^{12}{\rm CO})$ gas and
    star-forming regions) is offset from the peak of the line brightness
    toward lower velocities, i.e., the downstream side of molecular
    spiral arms.
    This implies that {\stateone} (warm gas) develops at the downstream
    side of the spiral arm and forms stars there.
\end{itemize}

These analyses are based on the pixel-by-pixel brightness distribution
and are free from the process of cloud identification.
Our method is advantageous in investigating the molecular gas content
in the Galaxy as a whole,
since the majority of the emission in our field of view is
found as the diffuse, more extended component ({\statetwo}),
which does not come under the traditional ``cloud'' classification.


\acknowledgments

The 45-m radio telescope is operated by NRO,
a branch of National Astronomical Observatory of Japan.
The ASTE project is driven by NRO, in collaboration with University of
Chile, and Japanese institutes including University of Tokyo,
Nagoya University, Osaka Prefecture University, Ibaraki University, and
Hokkaido University.
We are grateful to T.\ M.\ Dame for providing the
$^{12}{\rm CO}\; J=1\mbox{--}0$ data\-set taken with the CfA 1.2-m
telescope.
We thank J. Barrett for improving the manuscript.
A part of this study was financially supported 
by the MEXT Grant-in-Aid for Scientific Research on
Priority Areas No.\ 15071202.

{\it Facilities:} \facility{No:45m}, \facility{ASTE}.


\appendix

\section{Possible Deviation from the Assumed Structure of the Galaxy}
\label{app:sgrarm}

Although we assumed the structure of the Galaxy briefly described in
Section \ref{sec:field}, the precise picture is the topic under debate.
\citet{georgelin1976} referred to the Sgr and Sct arms as
``major'' and ``intermediate'' arms,
on the basis of a study of \ion{H}{2} regions.
``Warm'' or ``hot'' molecular clouds follow the $l$-$v$ loci
of the \ion{H}{2} regions in these arms
\citep{solomon1985,sanders1985,scoville1987}, as mentioned in
Section \ref{sec:introduction}.
These studies suggested that the Sgr arm is a major and prominent arm.
On the other hand, \citet{drimmel2000} presented that at the Sgr
tangent ($l\simeq 50\degr$) there is no enhancement in the
{\it K} band, as opposed to the Sct tangent ($l\simeq 30\degr$),
and suggested that the Sgr is an interarm or secondary arm structure.
Similar results were also achieved from mid-infrared star count
\citep{benjamin2005,churchwell2009}.

In this paper we regarded the radial velocities at which the
total CO intensity within the field of view takes its maxima
as ``spiral arm'' velocities, and found that they coincide with
the {\it traditional} Sgr arm velocities \citep[e.g.,][]{sanders1985}.
We then focused on the relationship between the indices of the
gas properties (BDF/BDI, line ratios, and \ion{H}{2} regions)
and the arm velocities.
Our study, therefore, relies on the existence of the Sgr arm and
the assumption that the observed maxima of the total CO intensity
correspond to the Sgr arm, but is independent of the details of the
structure of the Galaxy (i.e., whether the Sgr arm is a major or
secondary arm, and its exact location in the Galaxy).
The location (distance) of the observed gas content is affected by
the near-far distance ambiguity and the deviation from the flat,
circular rotation of the Galaxy.
The near-far ambiguity was taken into account in the analyses of
the BDI and BDF and the difference between the arm and the inter-arm
regions was proven to be firm.
The implication that the high-BDI gas and \ion{H}{2} regions
are located in the outer side of the spiral arm (from the fact that
they have lower radial velocities than the brightness peak)
is unchanged regardless of the near-far ambiguity,
unless the local non-circularity (streaming motion)
overrides the global trend of the velocity field of the Galaxy.

\section{Systematic Errors in Intensity Ratios}\label{app:efficiency}

The $T_{\rm MB}$ brightness scale is likely overestimated
because the sources are, in general, bigger than the main beam sizes of
the telescopes.
In particular, $\eta_{\rm MB}$ of the 45-m telescope differs
significantly from $\eta_{\rm moon}$, which means a considerable amount
of power comes from the side\-lobe.
Since the emission in the velocity range of 70--80 ${\rm km\;s^{-1}}$
is widespread over the field of view,
the coupling efficiency $\eta_{\rm c}$ should be close to
$\eta_{\rm moon}$ rather than $\eta_{\rm MB}$.
Therefore we adopt $\eta_{\rm c} \simeq \eta_{\rm moon} = 0.69$
for this component.
On the other hand, the high-brightness structures are less affected
because of their low surface filling factors.
The typical size of these structures is up to a few arc\-minutes
(Section \ref{subsec:chmap}).
The 45-m telescope, whose dish consists of 1- to 2-m reflection panels,
is considered to have an error patterns of $5\arcmin\mbox{--}10\arcmin$
width.
Thus a high-brightness structure fills only a small part of the error
pattern: we use $\eta_{\rm MB}$ as the lower limit of $\eta_{\rm c}$.
We consider that the error of the CO $J=3\mbox{--}2$ brightness
due to the error pattern of the ASTE telescope is small
compared with that of the $J=1\mbox{--}0$,
given the high $\eta_{\rm MB}$ of the telescope.



\clearpage

\clearpage

\clearpage

\begin{figure}
\epsscale{0.8}
\plotone{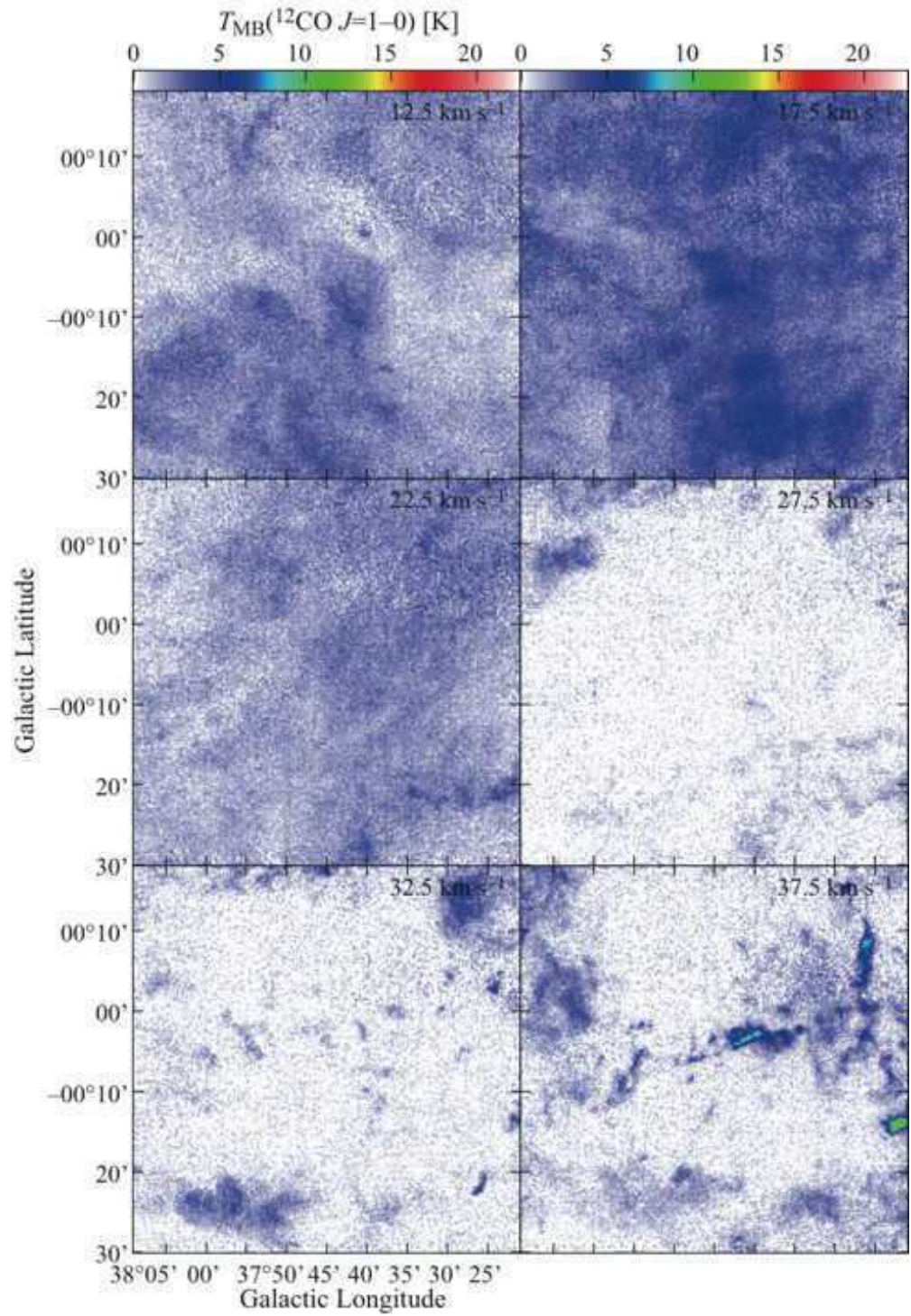}
\caption{Velocity channel maps of $^{12}{\rm CO}$ $J=1\mbox{--}0$
  $T_{\rm MB}$.
  The centroid velocity of each 5-${\rm km\; s^{-1}}$ channel is shown
  at the top-right corner.
  \label{fig:chmap12_1}}
\end{figure}

\clearpage

\begin{figure}
\addtocounter{figure}{-1}
\epsscale{0.8}
\plotone{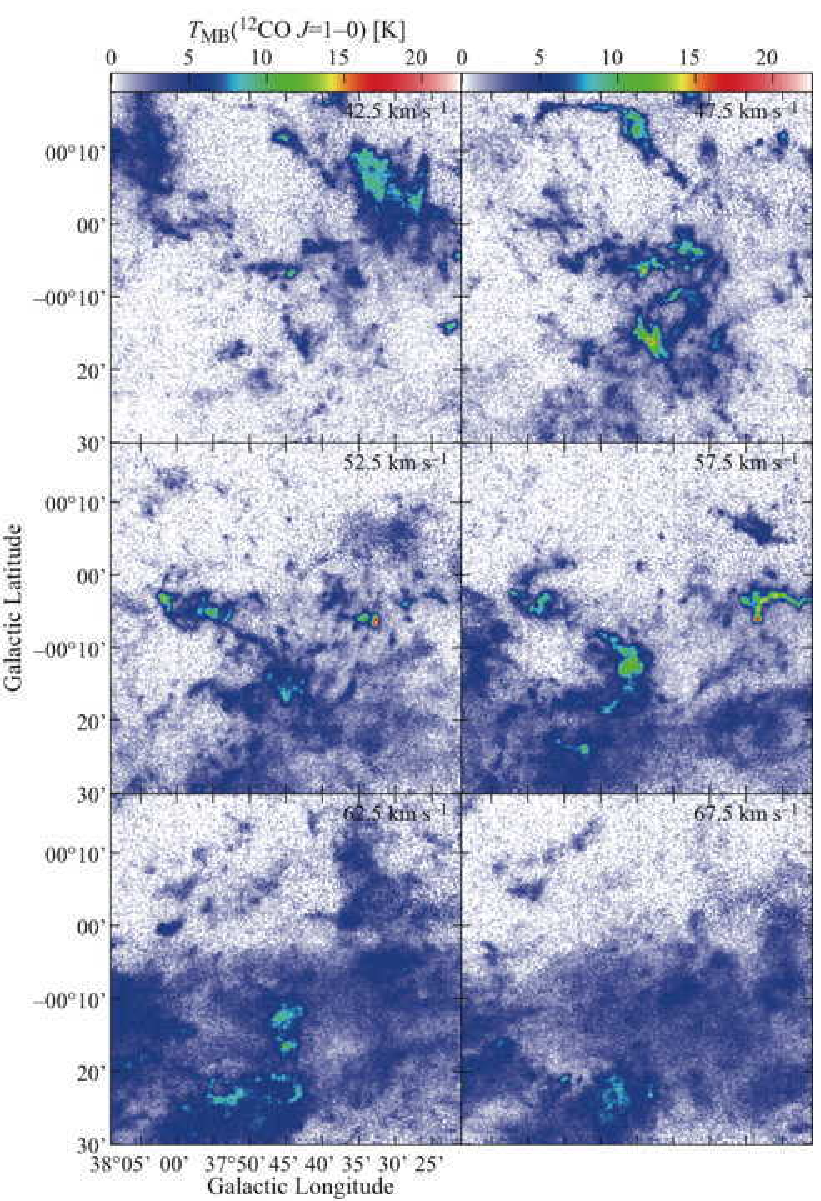}
\caption{{\it Continued}
  \label{fig:chmap12_2}}
\end{figure}

\clearpage

\begin{figure}
\addtocounter{figure}{-1}
\epsscale{0.8}
\plotone{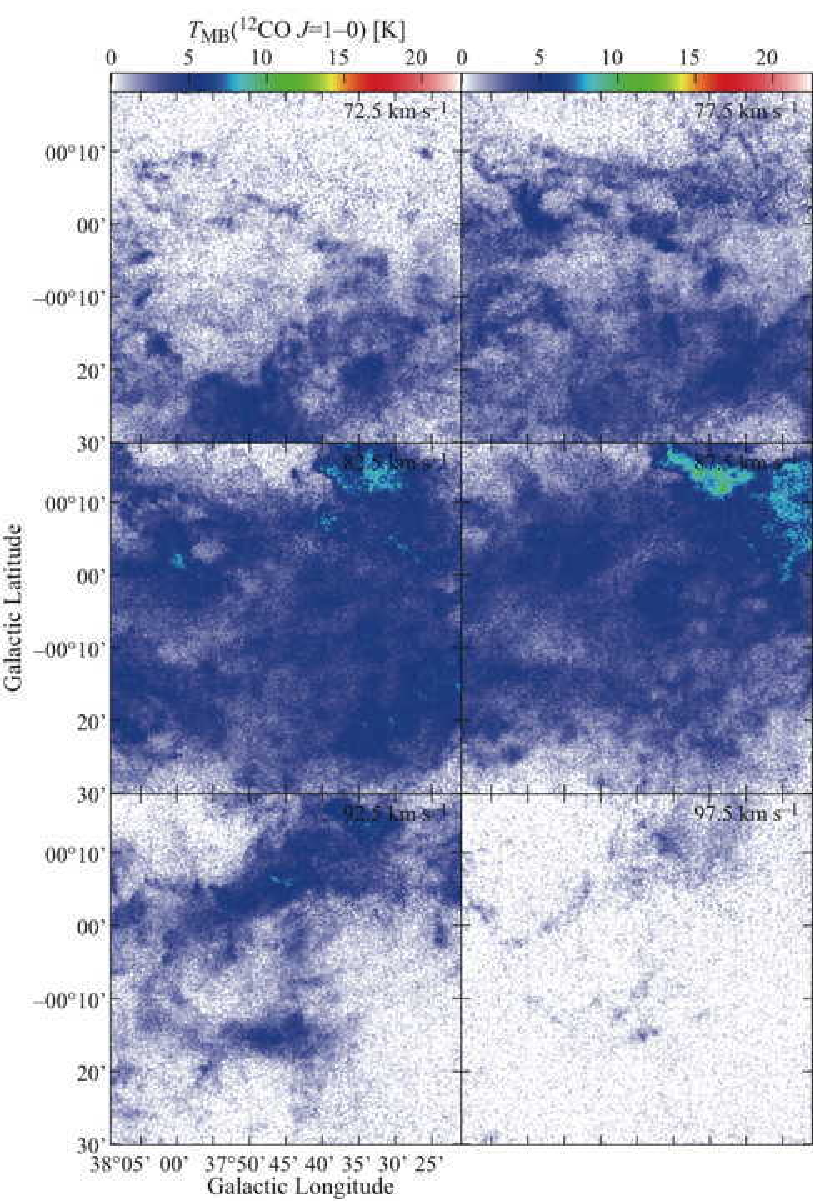}
\caption{{\it Continued}
  \label{fig:chmap12_3}}
\end{figure}

\clearpage

\begin{figure}
\epsscale{0.8}
\plotone{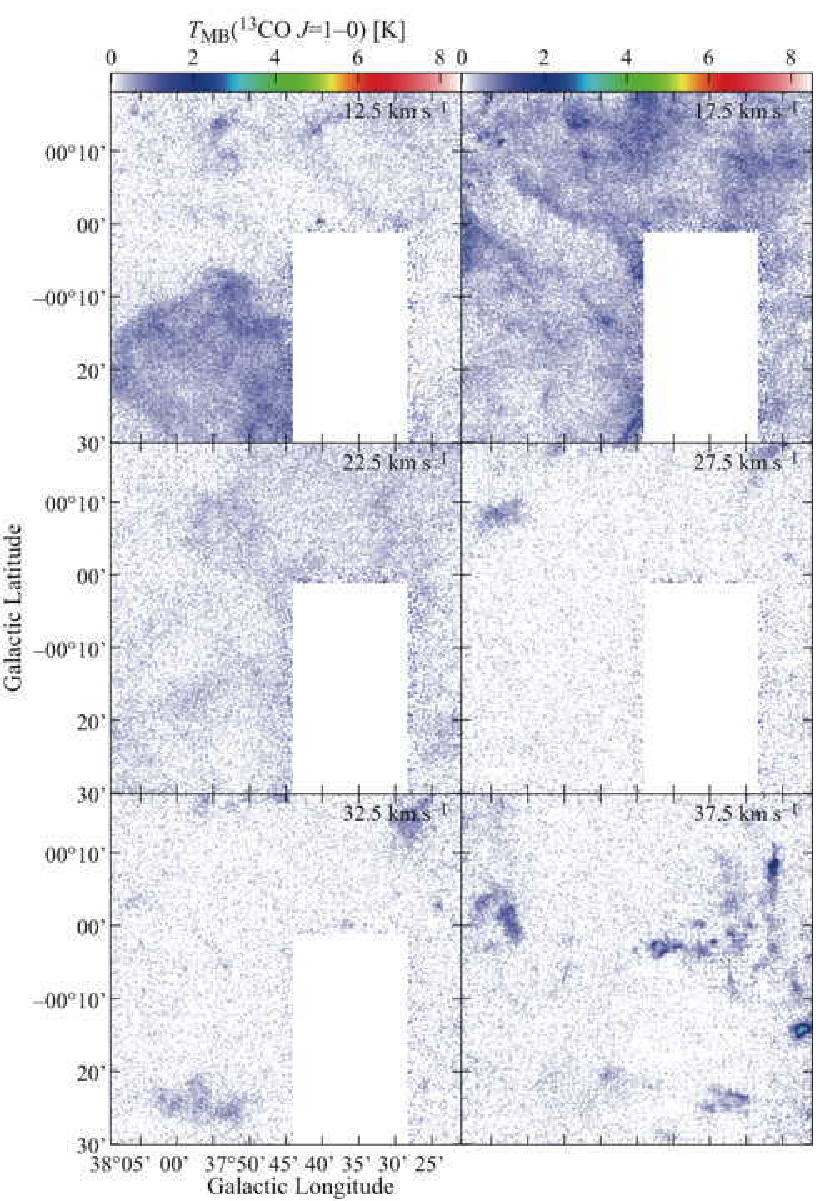}
\caption{Same as Fig.\ \ref{fig:chmap12_1}, but for the $^{13}{\rm CO}$
  $J=1\mbox{--}0$ line.
  \label{fig:chmap13_1}}
\end{figure}

\clearpage

\begin{figure}
\addtocounter{figure}{-1}
\epsscale{0.8}
\plotone{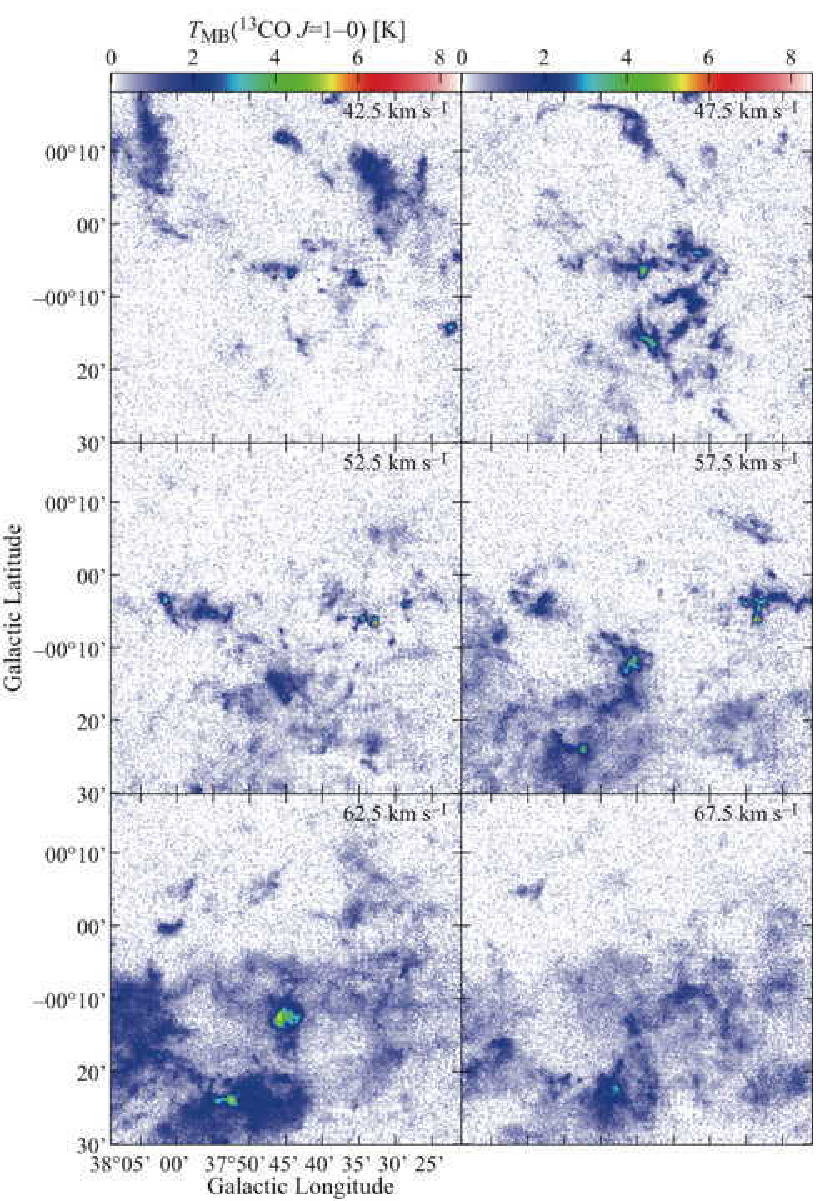}
\caption{{\it Continued}
  \label{fig:chmap13_2}}
\end{figure}

\clearpage

\begin{figure}
\addtocounter{figure}{-1}
\epsscale{0.8}
\plotone{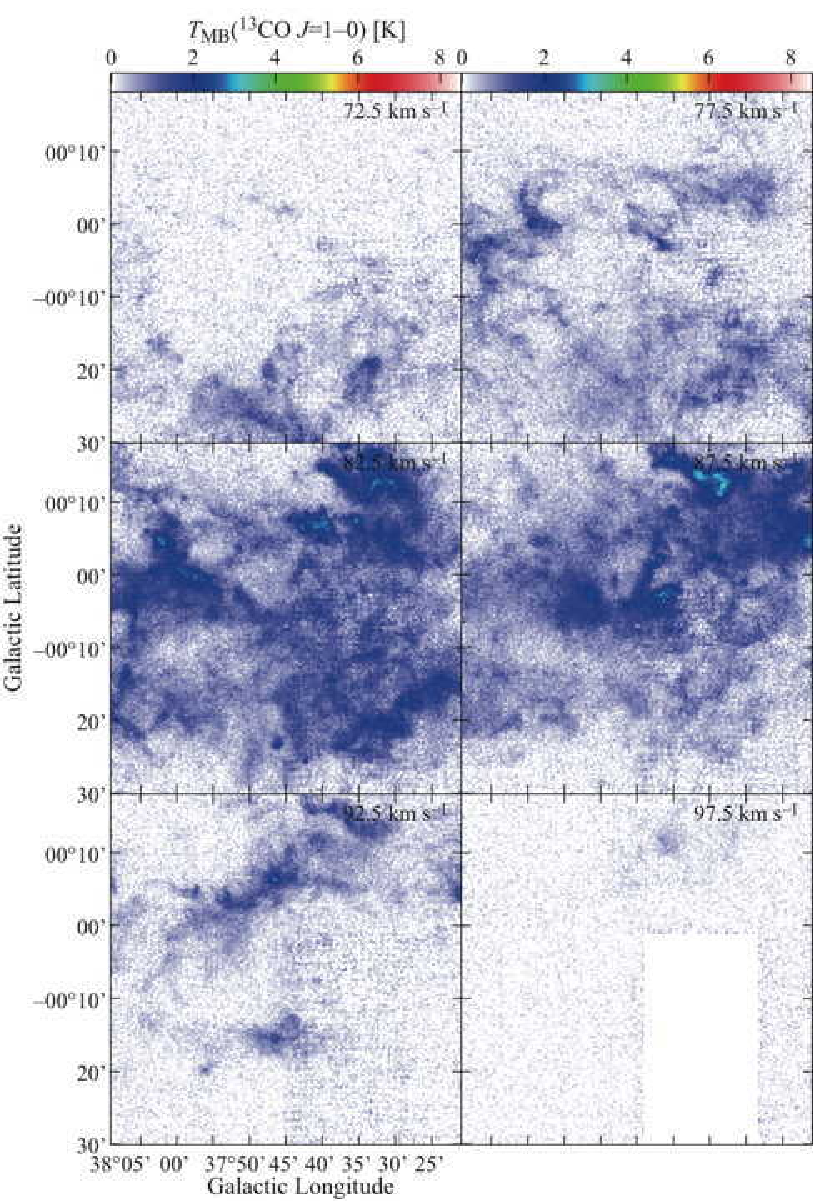}
\caption{{\it Continued}
  \label{fig:chmap13_3}}
\end{figure}

\clearpage

\clearpage

\clearpage

\clearpage

\clearpage

\clearpage

\clearpage

\clearpage

\clearpage

\begin{deluxetable}{lcccc}
\rotate
\tablecolumns{5}
\tablewidth{0pc}
\tablecaption{Parameters of the Observations and Reduced Maps
  \label{tab:obs}}
\tablehead{
  \colhead{} & \colhead{Period 1} & \colhead{} & \multicolumn{2}{c}{Period 2} \\
  \colhead{} & \colhead{(2002 Nov -- 2003 May)} & \colhead{} & \multicolumn{2}{c}{(2005 Dec -- 2006 Mar)} \\
  \cline{2-2} \cline{4-5} \\
  \colhead{} & \colhead{$^{13}{\rm CO}\; J=1\mbox{--}0$} & \colhead{} & \colhead{$^{12}{\rm CO}\; J=1\mbox{--}0$} & \colhead{$^{13}{\rm CO}\; J=1\mbox{--}0$}
}
\startdata
Grid spacing & $13\farcs 7$ (PSW) & & Nyquist (OTF) & Nyquist (OTF) \\
Area (${\mit\Delta}l \times {\mit\Delta}b$) & $0\fdg 3 \times 0\fdg 5$ & & $0\fdg 8 \times 0\fdg 8$ & $0\fdg 8 \times 0\fdg 8$\tablenotemark{a} \\
Bandwidth & 32 MHz (87.0 ${\rm km\;s^{-1}}$) & & 512 MHz (1330 ${\rm km\;s^{-1}}$) & 32 MHz (87.0 ${\rm km\;s^{-1}}$) \\
& \nodata & & \nodata & 512 MHz (1390 ${\rm km\;s^{-1}}$) \\
Frequency resolution & 37.8 kHz (0.10 ${\rm km\;s^{-1}}$) & & 1 MHz (2.6 ${\rm km\;s^{-1}}$) & 62.5 kHz (0.17 ${\rm km\;s^{-1}}$) \\
& \nodata & & \nodata & 1 MHz (2.7 ${\rm km\;s^{-1}}$) \\
Main-beam efficiency & $0.46\pm 0.03$ & & $0.39\pm 0.03$ & $0.45\pm 0.03$ \\
System noise temperature (DSB) & 300--600 K & & 350--450 K & 300--400 K \\
Map grid & $6\farcs 85$ & & $6\arcsec$ & $6\arcsec$ \\
Effective HPBW & $20\arcsec$ & & $17\arcsec$ & $17\arcsec$ \\
RMS noise ($T_{\rm MB}$) & 0.57 K (at 0.2 ${\rm km\;s^{-1}}$) & & 0.44 K (at 2.6 ${\rm km\;s^{-1}}$) & 0.62 K (at 0.2 ${\rm km\;s^{-1}}$) \\
& \nodata & & \nodata & 0.16 K (at 2.6 ${\rm km\;s^{-1}}$)
\enddata
\tablenotetext{a}{Except for the area observed in Period 1.}
\end{deluxetable}

\clearpage
\begin{deluxetable}{cccc}
\tablecolumns{4}
\tablewidth{0pc}
\tablecaption{\ion{H}{2} Regions in Our Field of View\label{tab:hii}}
\tablehead{
  \colhead{Name\tablenotemark{a}} & \colhead{$v_{\rm LSR}$} &
  \colhead{Near/Far} & \colhead{$d$}\\
  \colhead{} & \colhead{[${\rm km\;s^{-1}}$]} &
  \colhead{} & \colhead{[kpc]}
}
\startdata
U37.37$-$0.24 & 39.4 & F & 11.2\tablenotemark{b}\\
D37.37$-$0.07 & 53.2 & F & 10.2\\
D37.44$-$0.04 & 53.0 & F & 10.2\\
U37.55$-$0.11 & 48.9 & F & 10.5\\
C37.64$-$0.11 & 52.3 & F & 10.2\\
C37.67$+$0.13 & 88.9 & T & 6.7\\
U37.75$-$0.10 & 49.7 & F & 10.4\\
U37.76$-$0.20 & 65.8 & F & 9.3\\
U37.87$-$0.40 & 59.2 & F & 9.7\\
C38.05$-$0.04 & 58.3 & F & 9.7
\enddata
\tablenotetext{a}{Entries are taken from \citet{anderson2009a}.}
\tablenotetext{b}{The two methods to solve the near-far ambiguity
(the emission/absorption method and the self-absorption method)
disagree with each other.}

\end{deluxetable}

\end{document}